\documentclass[aps,prd,twocolumn,noshowpacs,superscriptaddress,groupedaddress,nofootinbib,preprintnumbers]{revtex4}
\usepackage{amsmath}
\usepackage{amsfonts}
\usepackage{amssymb}
\usepackage{textcomp}
\usepackage{graphicx, rotating}
\usepackage{epstopdf}
\usepackage{epsfig}
\usepackage{latexsym}
\usepackage{graphicx}
\usepackage{color}
\usepackage{slashed}
\usepackage{hyperref}
\usepackage{multirow}
\usepackage{subfigure}
\usepackage{epstopdf}
\usepackage{appendix}
\usepackage{comment}
\usepackage{bbold}
\usepackage{color}
\usepackage{setspace}
\usepackage{footnote}

\usepackage[]{units}

\begin{document}
\title{Neutrinos in Large Extra Dimensions and Short-Baseline $\nu_e$ Appearance}

\author{Marcela~Carena}
\affiliation{Theoretical Physics Department, Fermi National Accelerator Laboratory,
PO Box 500, Batavia, IL 60510, U.S.A.}
\affiliation{Enrico Fermi Institute, University of Chicago, Chicago, IL 60637, U.S.A.}
\affiliation{Kavli Institute for Cosmological Physics, University of Chicago, Chicago, IL 60637, U.S.A.}

\author{Ying-Ying~Li}
\affiliation{Department of Physics, The Hong Kong University of Science and Technology,
Clear Water Bay, Kowloon, Hong Kong S.A.R., P.R.C}

\author{Camila~S.~Machado}
\affiliation{Instituto de F\'isica Te\'{o}rica, Universidade Estadual Paulista, SP, Brazil}

\author{Pedro~A.~N.~Machado}
\affiliation{Theoretical Physics Department, Fermi National Accelerator Laboratory,
PO Box 500, Batavia, IL 60510, U.S.A.}

\author{Carlos~E.~M.~Wagner}
\affiliation{Enrico Fermi Institute, University of Chicago, Chicago, IL 60637, U.S.A.}
\affiliation{Kavli Institute for Cosmological Physics, University of Chicago, Chicago, IL 60637, U.S.A.}
\affiliation{High Energy Physics Division, Argonne National Laboratory, Argonne, IL 60439}

\date{\today}

\begin{abstract}
We show that, in the presence of bulk masses, sterile neutrinos propagating in large extra dimensions (LED) can induce electron-neutrino appearance effects. This is in
 contrast to what happens in the standard LED scenario and hence LED models with explicit bulk masses 
have the potential to  address the MiniBooNE and LSND appearance results, as well as the reactor and Gallium anomalies. A special feature in our scenario is that the mixing of the first KK modes to active neutrinos can be suppressed, making the contribution of heavier sterile neutrinos to oscillations relatively more important.
We study the implications of this  neutrino mass generation mechanism for current and future neutrino oscillation experiments, and show that the Short-Baseline Neutrino Program at Fermilab will be able to efficiently probe such a scenario. 
In addition, this framework
leads to massive Dirac neutrinos and thus precludes any signal in neutrinoless double beta decay experiments. 
\end{abstract}

\preprint{EFI-17-21, FERMILAB-PUB-17-338-T}

\maketitle


\section{Introduction}

Unambiguous measurements of neutrino  oscillations in the past two 
decades have provided clear evidence that neutrinos have non-vanishing 
masses and that the mass eigenstates are non-trivial admixtures of the flavor eigenstates. 
In fact, it is well understood that there are two small 
but quite different mass splittings, leading to flavor oscillations at macroscopic 
distances. For neutrino energies in the range of a few MeV  the smaller (``solar'') 
mass splitting $\Delta m^2_{21}$ induces  neutrino oscillations for baselines of few hundred kilometers, while the larger 
(``atmospheric'')  splitting $\Delta m^2_{31}$ would induce oscillations at baselines of about one kilometer.
Moreover, the Pontecorvo-Maki-Nakagawa-Sakata (PMNS) neutrino mixing  matrix~\cite{Pontecorvo:1957cp,Maki:1962mu} is found to have large 
off-diagonal entries, at variance with the quark Cabibbo-Kobayashi-Maskawa (CKM) mixing matrix that has only small off-diagonal entries.

Despite the well understood 3-neutrino paradigm, there are indications
of neutrino oscillations at very short baselines, that would call for additional mass splittings, beyond
the solar and atmospheric ones aforementioned. Perhaps the most intriguing is the one associated with $\bar\nu_\mu\to\bar\nu_e$  appearance at the LSND experiment~\cite{Athanassopoulos:1995iw,Athanassopoulos:1996jb,
Athanassopoulos:1997pv,Athanassopoulos:1997er, Aguilar:2001ty}, and its recent reincarnation at the MiniBooNE experiment~\cite{Aguilar-Arevalo:2012fmn,Aguilar-Arevalo:2013pmq}. MiniBooNE 
ran in both neutrino and antineutrino modes, and in each channel an excess was observed. In the neutrino mode, the excess was found mostly at low 
neutrino energies, below $475\,\unit{MeV}$, while in the antineutrino mode 
the excess events range from $200$ to about $1250\,\unit{MeV}$. If these anomalies were to be interpreted as neutrino oscillations, they would concurrently point to a much larger mass splitting $\Delta m^2\sim1\,\unit{eV}^2$, and an effective mixing angle $\sin^2 2\theta_{\mu e}=4|U_{e4}U_{\mu4}|^2\sim0.003$, where $U$ is the PMNS matrix with one additional sterile neutrino.

A different
anomaly, dubbed ``reactor antineutrino anomaly'', is associated with an apparent reduction of the flux of reactor electron anti-neutrinos with respect to its expected value~\cite{Mention:2011rk, Huber:2011wv}, something that
may be interpreted as neutrinos being converted into sterile neutrinos at short propagation distances. However, there has been some
observation of isotope dependence of this flux reduction~\cite{An:2017osx} that, if verified, will weaken the case for eV sterile neutrinos as an explanation of the reactor anti-neutrino anomaly.
On the other hand, there is a similar discrepancy between expected and observed
electron-neutrino events  in the calibration of Gallium experiments~\cite{Bahcall:1994bq,Giunti:2010zu,Anselmann:1994ar}.
Both Gallium and reactor anomalies, if interpreted via neutrino oscillations, point to  sterile neutrinos with  $\Delta m^2\sim 1$~eV$^2$ or higher and an effective mixing angle $\sin^2 2\theta_{ee}=4|U_{e4}|^2\sim0.1$.

Beyond these observational issues, the mechanism behind neutrino masses is still 
unknown. An interesting realization comes from models of flat large extra dimensions (LED),  in 
which right-handed neutrinos are allowed to propagate in the 
bulk of the extra dimensions, while the Standard Model (SM) fermions are restricted to live in the 4-dimensional  
brane~\cite{ArkaniHamed:1998vp, Dienes:1998sb, Dienes:2000ph,Dvali:1999cn, Barbieri:2000mg,Davoudiasl:2002fq, Machado:2011jt, Berryman:2016szd,Cao:2003yx, Mohapatra:2000wn}. The neutrino 
Yukawa couplings become tiny due to a volume suppression, leading to naturally light Dirac neutrinos. As a by-product of this type of models, a tower of Kaluza-Klein (KK) sterile neutrinos arises 
with masses proportional to  the inverse radius $R$ of the extra 
dimension. Only the lower mass states of the tower mix in a relevant way with the SM-like neutrinos.  
When $R\sim\mu m$, these states are  at the eV 
scale and thus the anomalies observed in short-baseline oscillation experiments 
could in principle be a consequence of this neutrino mass generation 
mechanism. 

Such mechanism in LED is quite appealing, and
can lead to neutrino disappearance from oscillations into
sterile neutrinos at short baseline experiments.
However, it cannot explain $\nu_\mu \to \nu_e$ appearance
effects~\cite{Davoudiasl:2002fq}, as suggested  by LSND and MiniBooNE data. 
Models with more sterile Dirac fermions or extra dimensions with different radii are proposed in Ref.~\cite{Davoudiasl:2002fq} for this. Adding Majorana mass terms~\cite{Diego:2008zu, Lukas:2000wn, Agashe:2000rw, Caldwell:2001dj,Lam:2001iy} may serve as an another direction to be fully explored. In this article, we shall
consider the possibility of adding Dirac bulk mass terms for the sterile neutrinos to accommodate neutrino appearance effects.
These bulk mass terms were introduced before, e.g. see Refs.~\cite{Diego:2008zu, Lukas:2000wn, Agashe:2000rw}, but here we show 
explicitly their effects on oscillations at short baselines, particularly for 
the $\nu_\mu\to\nu_e$ appearance mode.
It is worth mentioning  that the generation of neutrino masses in the deconstructed LED model with bulk mass terms is analogous to the clockwork mechanism for the obtention of the small neutrino Yukawa couplings~\cite{Hambye:2016qkf}.
Importantly, the LED scenario with bulk masses, that will be called here ``LED+'', leads to weak effects in the long-baseline neutrino experiments, but it can
be tested in the future Short-Baseline Neutrino Physics Program (SBN) at Fermilab~\cite{Antonello:2015lea}.
It would also lead to signatures at the Katrin Experiment~\cite{Angrik:2005ep}.

This article is organized as follows. In section \ref{LEDneutrinos} we define our framework. In section~\ref{fit} we concentrate on the 
phenomenology  of our model, evaluating the constraints from existing data and studying the possible explanations of the observed anomalies in neutrino oscillation experiments. We also estimate the impact of LED+ in future neutrino oscillation experiments. 
In section \ref{constraints}, we discuss the constraints from Higgs decays and cosmology. 
We reserve section \ref{conclusion} for our conclusions. 
In Appendix~\ref{fermion} we show an interesting correspondence of this model with the linear dilaton scenario while in Appendix~\ref{MFV} we present the details regarding the Minimal Flavor Violation (MFV) assumption that will be used in the analysis of the model. In Appendix~\ref{app:kinematics} we present details useful for the estimate of the constraints coming from kinematical tests of neutrino masses.

\section{Neutrinos in LED with Dirac bulk masses}
\label{LEDneutrinos}
 We consider  a $5$-dimensional flat space compactified on a $S_1/Z_2$ orbifold, with three generations of right-handed neutrinos propagating in the bulk,
and SM fermions restricted to the 4D brane. Regarding the SM singlets, it is 
more convenient to work on an ``intermediate'' mass basis in which the flavor mixing has already been diagonalized. In such basis, the kinetic and mass terms in the action are given by
\begin{equation}\label{eq:action-singlet}
S_f=\int d^4x\,dz\,\left[i\bar{\Psi}_i\Gamma^A \overset{\leftrightarrow}{\partial_{A}}\Psi_i- c_i\bar{\Psi}_i\Psi_i\right],
\end{equation}
with $\Gamma^A = (\gamma^\mu,  i \gamma^5)$, $z\in[0,\pi R]$ and $c_i$ 
being the bulk mass parameters. Note that lepton number is conserved in our Lagrangian, so no Majorana mass term is present. Here we will use $\alpha,\beta$ to denote flavor,
$i,j$ for the ``intermediate'' mass basis, and $n,m$ will be reserved to specify the KK-mode.
The 5D fermion $\Psi_i$ can be decomposed  as
\begin{equation}
\label{eq:decompose}
\Psi^{L,R}_i = \sum_{n}\psi^{L,R}_{i,n}(x) f^{L,R}_{i,n}(z),
\end{equation}
with $\Psi^{L,R}_i=P_{L,R}\Psi_i$ and $\gamma_5\Psi^{L,R}_i=\mp \Psi^{L,R}_i$. 
To have the Dirac action canonically normalized in four dimensions,
the wave functions $f^{L,R}_{i,n}(z)$ should satisfy the following normalization condition
\begin{eqnarray}
\label{eq:normalize}
\int^{\pi R}_{0} dz\, f^{L,R}_{i,n}(z) f^{L,R}_{i,m}(z) = \delta_{mn} .
\end{eqnarray}

The orbifold symmetry allows two choices of boundary conditions \cite{Ponton:2012bi}: either all left-chiral fields are odd functions (Dirichlet boundary conditions)
and all right-handed ones are
even functions, or vice-versa. 
In order to generate neutrino masses, there should be a right-handed chiral zero mode, therefore we will use the Dirichlet boundary conditions for the left-handed modes on both branes. Then, 
we get a right-handed massless zero mode with wave function
\begin{align}
f^R_{i,0}(z)=\sqrt{\frac{2c_i}{e^{2\pi Rc_i}-1}}e^{c_i z},
\label{eq:zero}
\end{align}
while for all other KK-modes we obtain
{\small{\begin{align}
&f^{L}_{i,n}(z)=\sqrt{\frac{2}{\pi R}}\sin \left(\frac{n z}{R} \right),\\
&f^{R}_{i,n}(z)=\sqrt{\frac{2}{\pi R \,m_{i,n}^2}}\left[c_i\sin \left(\frac{nz}{R} \right)+\frac{n}{R}\cos \left(\frac{nz}{R} \right)\right],\\ \label{eq:mass-i}
&(m^i_{n})^2 = \Big(\frac{n}{R}\Big)^2 + c_i^2 .
\end{align}}}

The bulk fermions will couple to SM neutrinos through 
the Yukawa terms in the IR brane~\cite{Grossman:1999ra}. In the ``intermediate'' basis, the Yukawa terms read
\begin{align}
\label{eq:Yukawa}
S_Y&=-\int d^4 x \sum_{i=1}^3\left(y_{i} \bar{L}_i \tilde{H} \Psi^{R}_i (x^{\mu}, 0) + \text{h.c.}\right)\\
&=-\int d^4 x   \sum_{i=1}^3\sum_{n=0}^\infty\left(Y^i_{n} \bar{L}_i \tilde{H}  \psi_{i,n}^{R}(x^{\mu})  + \text{h.c.}\right),\nonumber
\end{align}
with $\tilde{H} = i\sigma_2 H^\ast$ and the effective coupling $Y^i_{n} = y_{i} f^{R}_{i,n}(0)$. We define $y_{i} = \lambda^i/\sqrt{M_5}$ with $M_5$ the
fundamental scale of the extra dimensional theory and $\lambda^i$ being  free dimensionless parameters.  $M_5$ is related to the Planck scale by 
\begin{equation}
M_{\text{Pl}}^2 = M_5^{2+d} V_d,
\label{eq:Mpl}
\end{equation}
where $d$ is the number of extra dimensions and $V_d$ is their volume. 
Note that to have both the 5D Yukawa matrix and the singlet bulk mass matrix diagonalized in the ``intermediate'' basis, we have assumed  alignment between these two matrices that is equivalent to assume a Minimal Flavor Violation (MFV) scenario (see Appendix~\ref{MFV} for details). This assumption is not essential for this scenario to work, but it will greatly simplify the phenomenological analysis.
For simplicity, the Yukawas and the bulk masses in the intermediate basis  are taken to be real.
Therefore, there are no additional CP phases besides the standard $\delta_{CP}$  appearing in a 3-neutrino framework.

We define the relation between the flavor and ``intermediate'' bases as in \cite{Machado:2011jt}, namely 
\begin{equation}
  \nu_{\alpha,0}^L = U_{\alpha i}\,\nu_{i,0}^L,\quad 
  \Psi_{\alpha}=R_{\alpha i}\Psi_{i}.
  \label{eq:rotate1}
\end{equation}
In the above, $U_{\alpha i}$ is the PMNS matrix for the
standard three flavor neutrino model and $R$ is a matrix that diagonalizes the bulk masses and Yukawa couplings.
The mass matrix $M_i^{nm}$  in the intermediate basis reads
\begin{equation}
M_i =\begin{pmatrix}
v Y^i_0& 0 &\cdots& 0  \\
v Y^i_1&m^i_1&\cdots&0\\
\vdots &0&\ddots&0\\
v Y^i_n&0&\cdots&m^i_n
\end{pmatrix},
\end{equation}
where $v=174~\unit{GeV}$ is the Higgs vev.

For one extra dimension,  the Yukawa couplings of the zero and KK-modes are given by
\begin{align}
Y^i_0 &=\lambda^i\sqrt{\frac{2 }{M_5}} \sqrt{\frac{c_i}{e^{2c_i\pi R}-1}},\nonumber\\
Y^i_n &= \lambda^i\sqrt{\frac{2}{M_5 \pi R}}\sqrt{\frac{n^2}{n^2+ c_i^2 R^2}}.
\label{eq:yukawa-1}
\end{align}
We are interested in sterile neutrinos with masses of the order of 1~eV, implying at least one extra dimension with size $1/R={\cal{O}}$(1 eV).
If this were the only extra dimension, consistency with Eq.~(\ref{eq:Mpl}) would demand $M_5 \simeq 10^{10}$~GeV, and hence values of 
$\lambda_i = {\cal{O}}(10^{-4})$ would be necessary in order to obtain the correct active neutrino masses for $c_i = {\cal{O}}$(1/R).  
Alternatively, one can think of models in which neutrinos propagate in $d$ extra dimensions,  where the size of the additional extra dimensions  is much smaller than $R$, $R_{k>1}\ll R_1\sim R$. In such a case, under the assumption that the effects of the heavier KK modes from the extra dimensions with small radii $R_{k>1}$ can
be neglected, one obtains
\begin{align}
Y^i_0 &=\lambda^i\sqrt{\frac{2 }{M_5^d V_d}} \sqrt{\frac{c_i \pi R}{e^{2c_i\pi R}-1}}=\lambda^i\frac{M_5 }{M_{pl}} \sqrt{\frac{ 2 c_i \pi R}{e^{2c_i\pi R}-1}},\nonumber
\\ \label{eq:yuks}
Y^i_n &= \lambda^i\sqrt{\frac{2}{M_5^d V_d}}\sqrt{\frac{n^2}{n^2+ c_i^2 R^2}} = \lambda^i\frac{M_5}{M_{pl}}\sqrt{\frac{2 n^2}{n^2+ c_i^2 R^2}} ,
\end{align}
To derive the second equality in Eq.~(\ref{eq:yuks}) we have used Eq.~(\ref{eq:Mpl}).
We can now lower the value of $M_5$ in order to obtain values of $\lambda_i = {\cal{O}}(1)$. 
In our scenario with small bulk mass terms, for definiteness, we  fixed $M_5=10^{6}$~GeV. Note that with more than one extra dimension, we require not only Dirichlet boundary conditions for the left-handed modes, but also  that the derivative of the right-handed bulk fermion wavefunction with respect to coordinates in $R_{k>1}$ directions are zero at the boundaries. The boundary condition applied in this way would allow for only one zero mass mode with the wavefunction given above.

We define the left rotation that diagonalizes the mass matrix $M_i$ in the intermediate basis as
\begin{equation}
  W_i^{nn'}(M_i M_i^\dagger)^{n'm'}W_i^{mm'}=m^2_{i,n}\delta_{nm},
\end{equation}
where $m^2$ is a diagonal matrix in KK space. Thus, the final left rotation involving active neutrinos that diagonalizes the full mass matrix is given by
\begin{equation}
  \mathcal{U}_{\alpha i}^{0n} = U_{\alpha i}W_i^{0n}.
\end{equation}
Note that the other entries of $\mathcal{U}_{\alpha i}^{nm}$ are not observable, as the sterile neutrinos do not couple to the electroweak gauge bosons.
The oscillation amplitude among active neutrinos is given by
\begin{equation}\label{eq:amplitude}
  \mathcal{A}(\nu_{\alpha,0}\to\nu_{\beta,0}; L) = \sum_{i,n}\mathcal{U}_{\alpha i}^{0n}(\mathcal{U}_{\beta i}^{0n})^*\exp\left(i \frac{ m_{i,n}^2 L}{2E}\right),
\end{equation}
where $L$ is the experiment baseline, $E$ is the neutrino energy, and the superscripts indicating left-handedness have been dropped. 
In our numerical simulations, we shall include the matter effects by adopting a similar procedure as the one in Refs.~\cite{Barbieri:2000mg,Davoudiasl:2002fq,Machado:2011jt}, namely, we rotate the matter potential into the intermediate basis and include its effects in the diagonalization of the KK modes.

\begin{figure}
\centering
\includegraphics[width=0.45\textwidth]{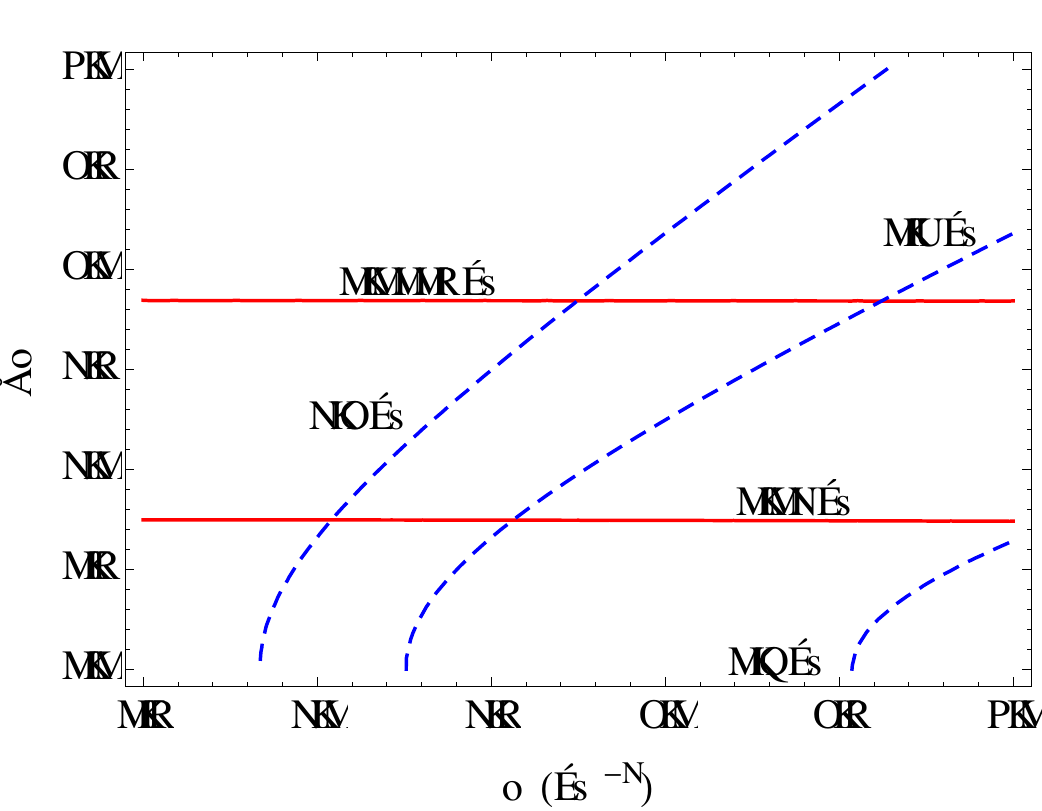}
\caption{The mass of the zero mode (red lines) and of the first KK mode (dashed-blue lines) as a function of the radius R and the bulk mass $c$ times $R$. We fix the scale $M_5 = 10^{6}~\text{GeV}$, $\lambda_i = 0.66$.
\label{fig:KK-modes}}
\end{figure}

Since the PMNS matrix has been fairly well constrained from neutrino oscillation experiments, the  effects of the KK modes can only be a perturbation over the standard three neutrino scenario. Therefore the $3\times 3$ block $\mathcal{U}_{\alpha i}^{00}$ would present a  slight deviation from unitary.
To connect this discussion more concretely with current neutrino data, first we note that the measured atmospheric and solar mass squared splittings correspond to $\Delta m^2_{\rm atm} \simeq m_{3,0}^2-m_{1,0}^2$ and $\Delta m^2_{\odot} \simeq m_{2,0}^2-m_{1,0}^2$, respectively. Moreover,
the observed approximate unitarity of the PMNS matrix~\cite{Parke:2015goa}, requires the deviations from unitarity to be at most about 10\%. That translates into a bound
\begin{equation}\label{eq:unitarity}
  \sum_{i=1}^3|\mathcal{U}_{\alpha i}^{00}|^2 \gtrsim 0.9
\end{equation}
for each flavor $\alpha$.
In Fig.~\ref{fig:KK-modes} we present isocontours of the masses of the 0-mode 
(red) and 1-mode (dashed blue) in the radius $R$ vs. $cR$ 
plane. The case $c=0$ is the LED scenario without bulk mass term, while 
$R\to0$ should recover Dirac neutrinos with the standard 3-neutrino framework. In the whole 
parameter space shown, the approximate bound from Eq.~(\ref{eq:unitarity}) is satisfied.

Based on Fig.~\ref{fig:KK-modes}, we define three benchmark points listed in Table~\ref{t:points} that will be used later on to perform a phenomenological study of the model. Point 1 realizes the normal ordering of the left-handed neutrinos, presenting relatively light KK modes with sizable mixing to the active neutrinos. In point 2, on the other hand, active neutrinos have inverted mass ordering, while the first KK mode masses are below 1~eV and the KK mixing is small. Finally, point 3 presents a degenerate neutrino spectrum (normal ordering) with KK modes around the eV scale and with large mixing to active neutrinos. A distinctive feature between LED without bulk masses and LED+ is the relation between KK mixing and the masses of active neutrinos and KK modes. 
In LED without bulk masses, the heavier  the active neutrino is, the larger is the mixing with KK modes, as explicitly shown later on in Eq.~(\ref{eq:mixing-LED-0}). Moreover, the first KK modes in each tower necessarily have the larger mixings  with the active neutrinos (as exemplified in Eq.~(\ref{eq:mixing-LED-0}) and by the crosses in Fig.~\ref{fig:kk-modes-pattern}).
In LED+ instead, the presence of non-zero $c_i$'s can dramatically change the above behaviors.
To exemplify the first, we present point 2 in Table~\ref{t:points}, where the lightest active neutrino has the largest mixing with the KK modes. For the latter, the turquoise circles in Fig.~\ref{fig:kk-modes-pattern} demonstrate that a non-zero $c_i$ can significantly suppress the mixing between active neutrinos and the first KK modes.

\begin{table}
\begin{center}
\begin{tabular}{l*{7}{c}r}
\hline
\rule{0pt}{4.5ex}   
 $\{P_a,\,\nu_i$\}            & $\dfrac{R}{{\rm eV}^{-1}}$ &~$c_i\,R$~~ &~~$\lambda^i$~~& ~$\dfrac{m^2_{i,0}}{{\rm eV}^2}$ & ~$\dfrac{m^2_{i,n^\prime}}{{\rm eV}^2}$ & $|W^{0n'}_i|^2$ \\ 
[1.5ex]\hline\hline 
\rule{0pt}{3ex} 
\{$P_1$, $\nu_1 $\} & $1.9$ & $4.24$ &$0.42$ & $\approx 0$ & $9.3$ & $9.0 \cdot 10^{-5}$ \\
~\{$P_1$, $\nu_2 $\}& $1.9$ & $1.19$ &$2.0$ & $7.6\cdot 10^{-5}$ & $0.66$ &  $0.0196$ \\
~\{$P_1$, $\nu_3 $\} & $1.9$ & $-0.037$ &$0.66$ & $2.5\cdot 10^{-3}$ & $0.27$ & $0.0169$ \\
\hline
\rule{0pt}{3ex} 
\{$P_2$, $\nu_1 $\} & $6.4$ & $-1.1$ &$0.27$ & $2.5\cdot 10^{-3}$ & $0.056$ & $5.9\cdot 10^{-3}$ \\
~\{$P_2$, $\nu_2 $\} & $6.4$ & $-1.2$ &$0.25$ & $2.6\cdot 10^{-3}$ & $0.066$ & $3.8\cdot 10^{-3}$\\
~\{$P_2$, $\nu_3 $\} & $6.4$ & $3.2$ &$1.1$ & $\approx 0$ & $0.64$ & $0.01$\\
\hline
\rule{0pt}{3ex} 
\{$P_3$, $\nu_1 $\} & $1.8$ & $0.43$ &$0.42$ & $1.9\cdot 10^{-4}$ & $0.37$ & $4.4 \cdot 10^{-3}$\\
~\{$P_3$, $\nu_2 $\} & $1.8$ & $1.0$ &$2.4$ & $2.6\cdot 10^{-4}$ & $0.65$ & $0.0361$\\
~\{$P_3$, $\nu_3 $\} & $1.8$ & $0.41$ &$1.7$ & $2.7\cdot 10^{-3}$ & $0.37$ & $0.0576$\\
\hline
\end{tabular} 
\end{center}
\caption{\label{t:points}Benchmark points used in the simulation, where $P_a$ for $a=\{1,2,3\}$ means point 1, 2, 3. The index $i$ is the ``intermediate basis'' index, we fixed $M_5= 10^{6}~\unit{GeV}$ and the index $n^\prime$ represents the KK mode that has the largest mixing with active neutrinos. }
\end{table}

\section{Neutrino  phenomenology}
\label{fit}
The phenomenology of neutrinos in large extra dimensions was 
widely studied for models without bulk mass terms (see e.g.\cite{Davoudiasl:2002fq, Machado:2011jt,Machado:2011kt,BastoGonzalez:2012me,Berryman:2016szd,Adamson:2016yvy}). In these realizations, the 
most striking signature is the disappearance of active neutrinos in short 
baseline oscillation experiments with a very regular pattern of masses and mixings. The appearance mode  is however absent in such studies precisely due to the regular behaviour of 
the KK spectrum and the structure of the mixing angles. In particular, there is sizable 
mixing  among flavors of the same KK mode (say, ``horizontally''), 
or among different KK modes of the same flavor (``vertically''). ``Diagonal'' mixing, that is, between different KK modes of different flavors is practically absent. Moreover, the ``horizontal'' mass splittings are always close to the atmospheric or solar mass splittings, and thus cannot mediate, e.g., $\nu_\mu\to\nu_e$ appearance at short baselines as suggested by the LSND and MiniBooNE anomalies.

To see this explicitly, we calculate the expression for Eq.~(\ref{eq:amplitude}) in the limit $c_i=0$. Using Eq.~(\ref{eq:mass-i}) and the approximation
\begin{align}\label{eq:mixing-LED-0}
W^{0n}_i\sim \frac{v Y^i_n}{m^i_n}\sim\frac{\sqrt{2} m_{i, 0}}{(n/R)},
\end{align}
for $n>0$, and $W_i^{00}\sim1$, we have
\begin{align}\label{eq:amplitude1}
  &\mathcal{A}(\nu_{\alpha,0}\to\nu_{\beta,0}; L) 
  \sim \sum_{i=1}^3U_{\alpha i}U_{\beta i}^* \exp\left(i\frac{\Delta m^2_{i,0} L}{2E}\right)  \nonumber\\
 & +2\sum_{i=1}^3\sum_{n=1}^\infty U_{\alpha i}U_{\beta i}^*\frac{m_{1,0}^2}{(n/R)^2}\exp\left(i \frac{n^2 L}{2R^2 E}\right) \nonumber \\
 &+2\sum_{i=2,3}\sum_{n=1}^\infty U_{\alpha i}U_{\beta i}^* \frac{\Delta m^2_{i,0}}{(n/R)^2}\exp\left(i \frac{ n^2 L}{2R^2E}\right)
\end{align}
with $\Delta m^2_{i,0} = m^2_{i, 0} - m^2_{1, 0}$ for $i=2,3$ to be the solar and atmospheric mass splitting, respectively. The first term in this approximation gives the standard 3-active neutrino oscillation result. For the appearance mode, $\alpha\neq\beta$, the second term vanishes, due to the unitarity of the PMNS matrix, and the third term contribution is suppressed by $\Delta m_{i,0}^2 R^2$, as  pointed out in Ref.~\cite{Davoudiasl:2002fq}. 

\begin{figure}
\centering
\includegraphics[width=0.45\textwidth]{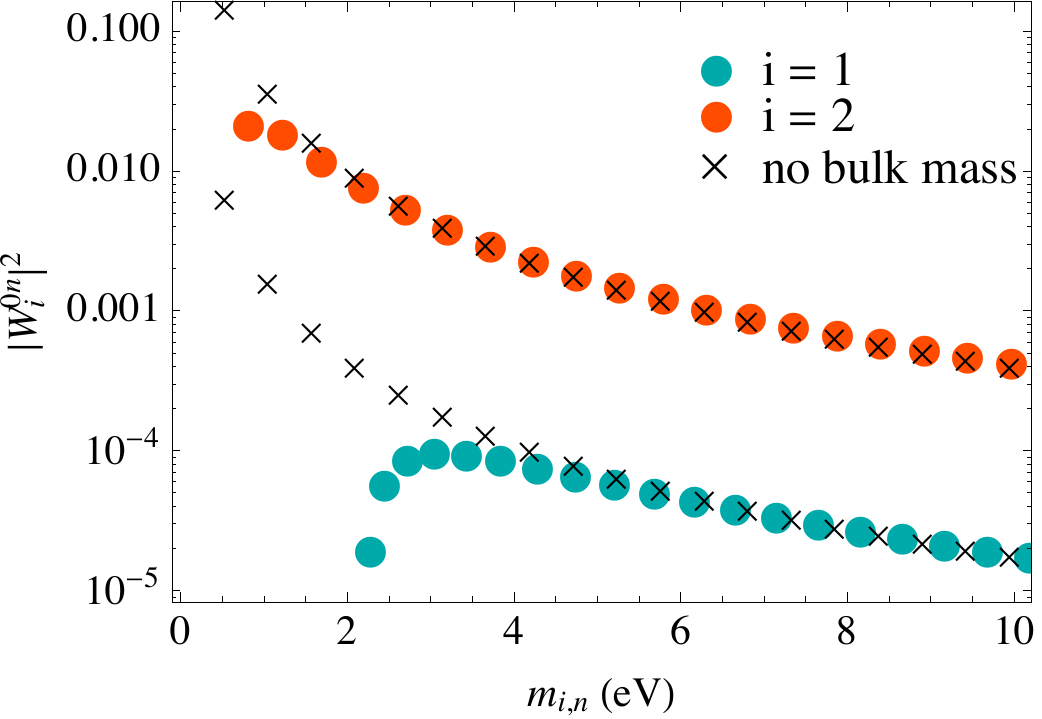}
\caption{\label{fig:kk-modes-pattern}Pattern of KK masses and mixings for the benchmark point 1 in Table~\ref{t:points} with $i=1$ (turquoise) and $i=2$ (red). For comparison, we also present the equivalent pattern for LED without bulk mass terms (crosses).}
\end{figure}

\begin{figure}
\centering
\includegraphics[width=0.45\textwidth]{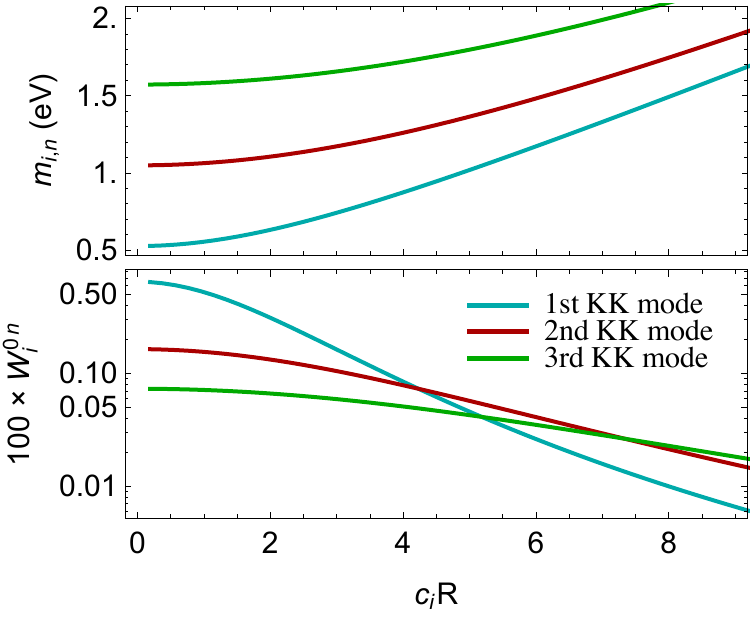}
\caption{\label{fig:impact-of-ci} Masses and mixings with active neutrinos of the first three KK modes as a function of the bulk mass parameter times the radius of the extra dimension, $c_i R$. The other LED+ parameters are taken to be $R=1.9$~eV$^{-1}$, $\lambda_i = 2.0$ and $M_5 = 10^6$~GeV.}
\end{figure}

The presence of bulk mass terms leads to a qualitatively different picture. As 
can be seen in Eqs.~(\ref{eq:mass-i}) and (\ref{eq:yukawa-1}), a non-zero $c_i$ will perturb the regularity of the 
masses in the corresponding KK tower, therefore enlarging the ``horizontal'' 
splittings for $n\ge1$. To exemplify this effect we show in Fig.~\ref{fig:kk-modes-pattern} the masses and mixings $|W^{0n}_i|^2$ between neutrino $\nu_{i,0}$ and the $n^{\text{th}}$ KK-mode for the benchmark point 1 (see Table~\ref{t:points}) and $i=1,2$.
Moreover, in Fig.~\ref{fig:impact-of-ci} we show how the masses of the first three KK modes and their mixings with $\nu_{i,0}$ change as a function of $c_i R$. Notice that not only the KK mode mass but also the mixing with active neutrinos changes drastically for different values of the bulk masses.
As it is shown in Fig.~\ref{fig:impact-of-ci}, for increasing values of the  bulk masses,  the mixing with the first KK modes can be suppressed, thus enhancing the relative importance of the heavier modes. 

\begin{figure}
\centering
\includegraphics[width=0.45\textwidth]{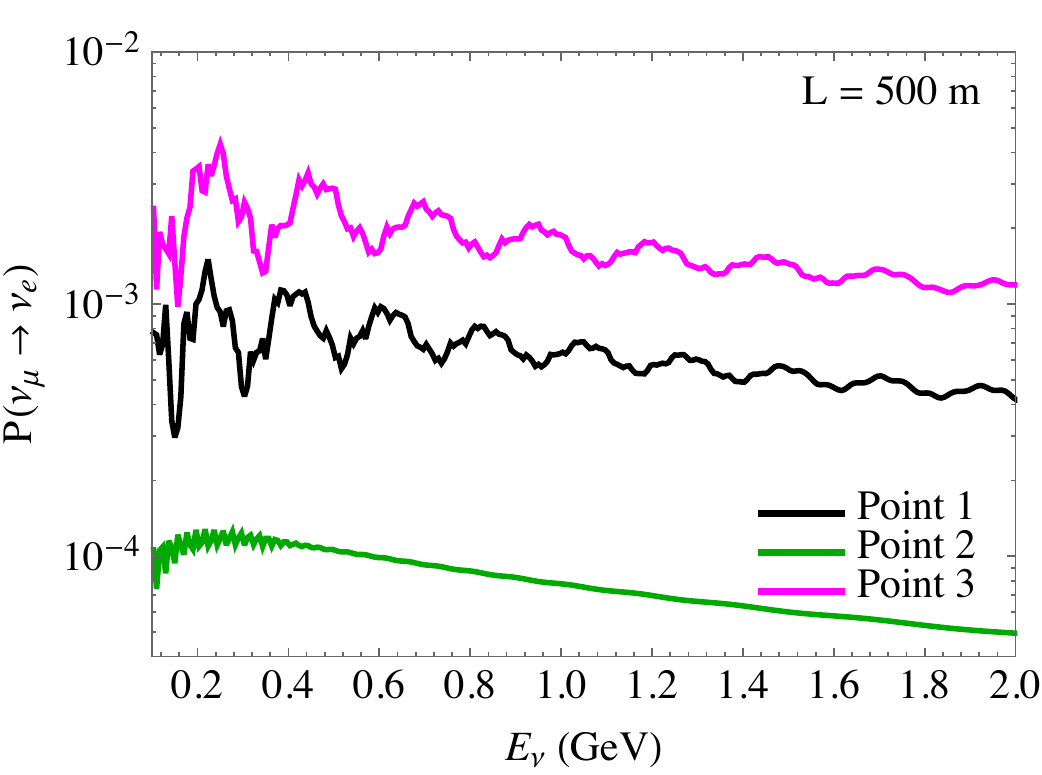}
\includegraphics[width=0.42\textwidth]{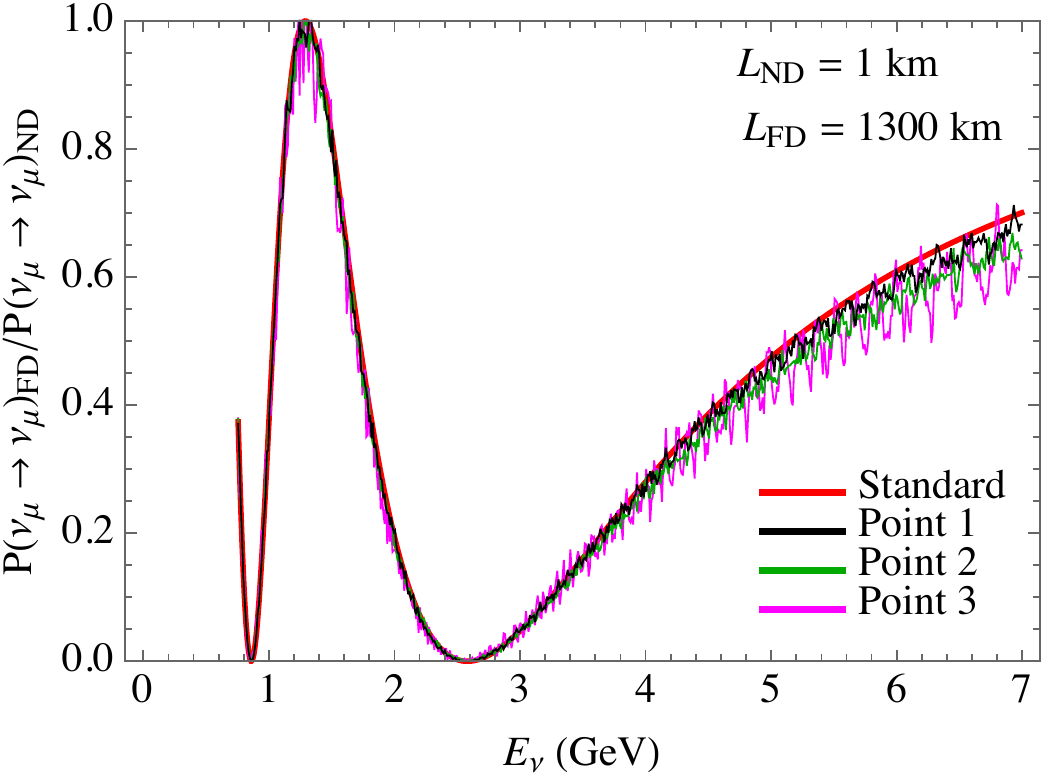}
\caption{\label{fig:probs} Oscillation probabilities for the three benchmark points given in Table~\ref{t:points} for short and long baselines as indicated in the plots. In the lower panel, the ratio of far-to-near detector probabilities  is presented.}
\end{figure}

The upper panel of Fig.~\ref{fig:probs} shows the oscillation probabilities at short baselines for the three benchmark points given in Table~\ref{t:points}. As can be clearly seen from Fig.~\ref{fig:probs}, bulk mass terms can lead to appearance at short-baseline neutrino experiments, possibly providing an explanation for the LSND and MiniBooNE anomalies.

We also expect that LED+ scenarios may have an impact in long-baseline oscillation experiments. 
The best way to look for heavy KK mode effects in long-baseline experiments would be in $\nu_\mu$ disappearance, as $\nu_\mu\to\nu_e$ appearance is suppressed by $\theta_{13}$ and also depends on $\delta_{CP}$ and the mass ordering. 
The LED+ effects on disappearance experiments yield fast oscillations that translate into an overall normalization change.  
For heavier KK modes, oscillations will also happen at the near detector and their effect will partially cancel in the near-to-far ratio~\cite{Bhattacharya:2011ee}.  In the lower panel of Fig.~\ref{fig:probs} we illustrate the $\nu_\mu$ disappearance effects in long-baseline experiments by showing the ratio of $\nu_\mu\to\nu_\mu$ oscillation probabilities between the near and far detectors. Notice that point~1 leads to a smaller effect due to the fact that active neutrinos mix with less KK modes compared to point 2 because of different values of $c_i$ and has a smaller mixing with a single KK mode compared to point 3.

We would also like to point out two important effects that should be taken into account when calculating the oscillation probabilities. First, the absolute values of $\Delta m^2_{31}$ should be slightly different between normal (points 1 and 3) and inverted hierarchy (point 2) cases in order to get the minimum of the oscillation at the same energy. This is due to the fact that the quantity that is measured in the $\nu_\mu\to\nu_\mu$ channel is the so-called $|\Delta m^2_{\mu\mu}|$, which is a function of the atmospheric and solar splittings, as well as the PMNS mixing angles. See Ref.~\cite{Nunokawa:2005nx} for detailed explanations. Second, to obtain percent level precision, for all benchmark points chosen, one needs to consider about 20 KK modes. This contrasts strongly with the LED scenario without bulk mass terms, in which 5 or 6 KK modes are enough to get to percent level precision calculations.

Below, we will analyze the current constraints coming from MINOS/MINOS+, NO$\nu$A, T2K, short baseline reactor experiments, LSND and MiniBooNE, as well as the future sensitivity of DUNE and the Short Baseline Neutrino Program at Fermilab. In principle, the IceCube experiment could also set a strong constraint on sterile neutrino models~\cite{TheIceCube:2016oqi} via a MSW resonance that would enhance active to sterile conversion when neutrinos cross the core of the Earth~\cite{Nunokawa:2003ep}. Nevertheless, we do not consider the IceCube sensitivity here for the following reason. 
As can be seen in Ref.~\cite{MINOS-website}, IceCube and MINOS/MINOS+ have comparable sensitivities to constrain sterile neutrinos with $\Delta m^2\sim 1~$eV$^2$. 
Nevertheless, for values of $\Delta m^2$ larger than $\sim 1~$eV$^2$, the sensitivity of IceCube degrades quickly, as the MSW resonance moves to higher energies for which the flux of atmospheric neutrinos becomes smaller.  Since in our scenario it is typical that many KK modes above 1~eV have sizeable contributions to the oscillation probability, we expect the IceCube bound to be weaker than MINOS/MINOS+.
Therefore, in this work we shall concentrate on accelerator and reactor oscillation experiments only.

\subsection{Past and present oscillation experiments}
\subsubsection{MINOS and MINOS+}
The standard LED scenario can be probed at long-baseline neutrino oscillation experiments.
The recent MINOS analysis~\cite{Adamson:2016yvy} using data collected from 2005 to 2012 excludes large extra dimension
 models with $R > (\unit[0.5]{eV})^{-1}$ at $90\%$ C.L., for a massless lightest neutrino. In addition, it is expected that MINOS+ will have a similar sensitivity to probe large extra dimension models~\cite{MINOS-website}.  As explainied above, 
the LED+ effects on disappearance experiments yield fast oscillations that translate into an overall normalization change.  
 The experimental sensitivity
will therefore be limited by  the overall normalization uncertainty which is about 5\%~\cite{Adamson:2016yvy}. 

\begin{figure}
\centering
\includegraphics[width=0.45\textwidth]{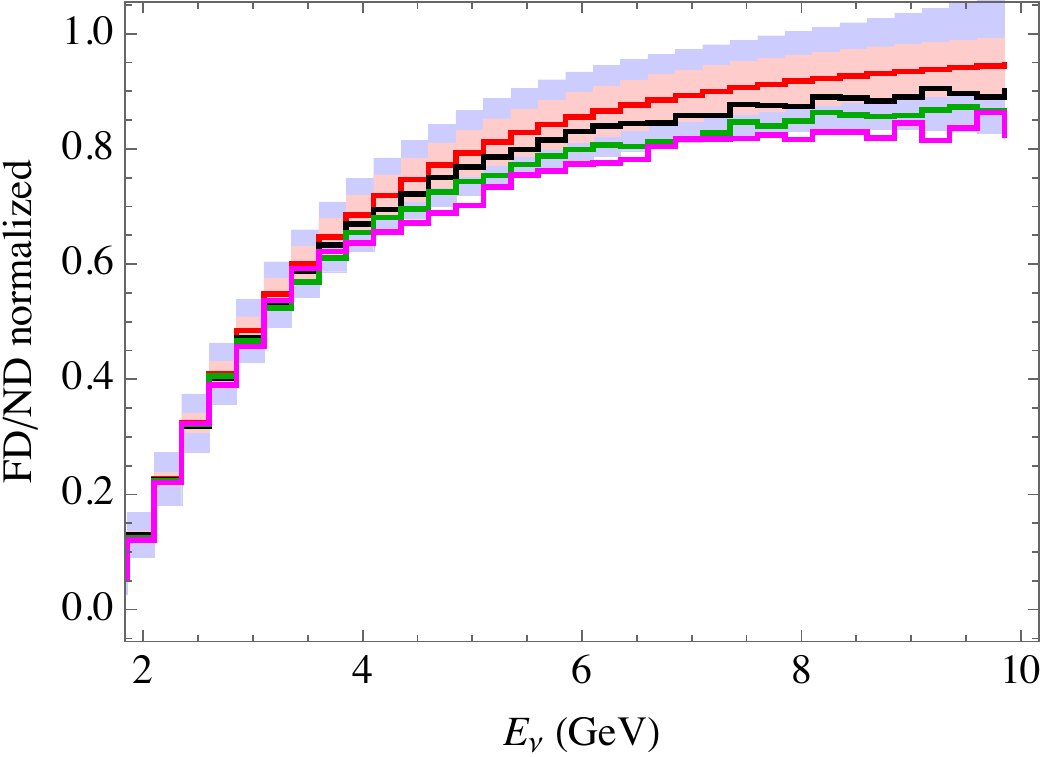}
\caption{Near-to-far ratio of events for the $\nu_\mu\to\nu_\mu$ disappearance channel at MINOS and MINOS+, normalized to 1 in the absence of oscillations. The red line is the expected ratio for the standard 3-neutrino framework. The light red  band shows a 5\% systematic uncertainty, while the blue band corresponds to the statistical uncertainty
 assuming full run~\cite{Adamson:2016yvy}. Black, green and magenta are the ratios for points 1,  2 and 3, respectively. \label{fig:minos}}
\end{figure}

To estimate the MINOS and MINOS+ sensitivity to LED+ models, we consider the combined flux, assuming $10.71\times 10^{20}$ and $5.8\times 10^{20}$ POT for the low and high energy beam configurations, respectively. The energy resolution and efficiency were taken from~\cite{Vahle-private}. In Fig.~\ref{fig:minos}, we illustrate the LED+ effects for the three benchmark points (Table~\ref{t:points}) which have $R^{-1}$ between $1.4-6.5~\unit{eV}$. 
We show the  near-to-far $\nu_\mu\to\nu_\mu$ ratios (normalized to 1 in the absence of oscillations) together with a 5\% normalization uncertainty (light red band) and the corresponding statistical uncertainty (light blue band) assuming full run~\cite{Adamson:2016yvy}. Note that the normalization uncertainty is fully correlated among energy bins, so it only applies to smeared fast oscillations, which is the case for the three benchmark points.
The benchmark points 1, 2, and 3 are depicted as the black, green and magenta lines, respectively. The first two are consistent within errors with the standard three flavor neutrino prediction (red line), while point 3 is marginally consistent (see also Fig.~\ref{fig:bound} later on).

\subsubsection{NO$\nu$\!A and T2K}
The current NO$\nu$A and T2K experiments may also constrain large extra dimension scenarios. Their sensitivity to LED without bulk masses was estimated in Ref.~\cite{Machado:2011jt}: no  improvement over MINOS sensitivity would be achieved. The reason is the following. 
The effects of KK mode oscillations in the $\nu_\mu\to\nu_\mu$ channel are more sizable at higher energies away from the atmospheric minimum.
However, NO$\nu$A and T2K are narrow band beam experiments, having the neutrino spectrum very localized at the atmospheric minimum.
Although this improves the sensitivity to the standard 3-neutrino oscillation parameters, it degrades the sensitivity to LED.

\begin{figure}
\centering
\includegraphics[width=0.45\textwidth]{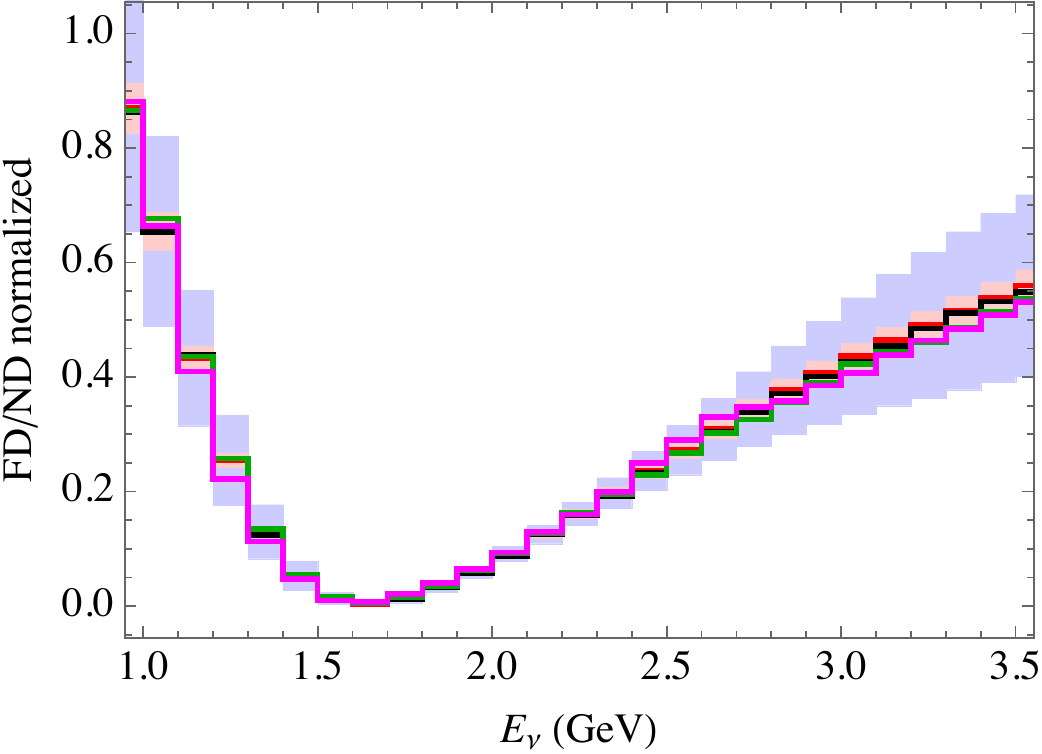}
\caption{Near-to-far ratio of events for the $\nu_\mu\to\nu_\mu$ disappearance channel at NO$\nu$A, normalized to 1 in the absence of oscillations. The red line is the expected ratio for the standard 3-neutrino framework. 
The light red  band shows a 5\% systematic uncertainty, while the blue band corresponds to the statistical uncertainty
 assuming $3\times10^{21}$ protons on target. Black, green and magenta are the ratios for points 1,  2 and 3 in Table~\ref{t:points}, respectively. \label{fig:nova}}
\end{figure}

To exemplify the impact of LED+ in these experiments, we present in Fig.~\ref{fig:nova} the near-to-far ratio of events in NO$\nu$A (normalized to 1 in the absence of oscillations)  for $3\times10^{21}$ protons on target, and 14 kton fiducial mass~\cite{Messier:1999kj, Paschos:2001np, Ambats:2004js, Yang_2004} for the
 three benchmark points. Points 1, 2, and 3 are depicted by the blue, green and magenta histograms, respectively, while  the red line corresponds to the standard 3-neutrino scenario. 
Both systematic (5\%) and statistical uncertainties are shown as the red and blue bands, respectively. Notice that Fig.~\ref{fig:nova} only depicts
neutrino energies lower than about 3.5~GeV, since above those energies the statistical error is fairly large due to the narrow band beam that peaks at the atmospheric oscillation minimum, as discussed above. 
Large deviations from the standard neutrino oscillation scenario happen at high energies, see Fig.~\ref{fig:probs}, not shown in Fig.~\ref{fig:nova}, and hence NO$\nu$A has limited sensitivity to test  our benchmark LED+ scenarios.  
One could wonder about what happens to the appearance channel, since LED+  may induce non-negligible $\nu_\mu\to\nu_e$ transitions. However, the appearance channel has low statistics and a strong dependence on $\theta_{23}$, $\delta_{CP}$ and the mass ordering, hence it is not expected to put any competitive bound on LED+.
The same features are present in T2K, and therefore we do not expect either of these two experiments to substantially improve MINOS and MINOS+ sensitivities to the LED+ scenario.

\subsubsection{Reactor experiments and the Gallium anomaly}
The reactor antineutrino anomaly is a discrepancy  between observed and predicted reactor antineutrino fluxes. At present, based on Refs.~\cite{Mention:2011rk, Huber:2011wv}, the measured neutrino flux at short baseline reactor experiments is  6\% below the theoretical flux prediction, with an associated uncertainty of  about 2\%. 
Recently, the Daya Bay analysis on the flux isotope dependence has shown that most of this discrepancy comes from the $^{235}$U
 isotope~\cite{Giunti:2016elf, An:2017osx}. Besides, other authors have proposed the use of  a larger, more conservative theoretical uncertainty of $5\%$, based on considerations of nuclear effects~\cite{Hayes:2016qnu}.
While this challenges the theory prediction for the fluxes and its associated 
uncertainties, the solution to the reactor anomaly puzzle is still far from clear. Here, we adopt an agnostic perspective and show the sensitivity of short baseline reactor neutrino experiments to LED+.

As it has been shown in Ref.~\cite{Machado:2011kt}, the reactor anomaly could in principle be explained by LED models via $\bar\nu_e$ mixing with KK modes. Similarly, this could also be explained in LED+ models.
We present in Fig.~\ref{fig:reactor} the predicted ratio of events between our scenario and no oscillations (as is the case for the standard 3 neutrino framework) for the three benchmark points in Table~\ref{t:points} and a representative baseline of $L = \unit[10]{m}$. The ratio between the total number of observed to expected events is depicted by the orange dashed line, with a 5\% associated theoretical uncertainty (light red band). Point 1 leads to  5\% disappearance ratio, point 2 allows for about 3\%,  while point 3 shows around 10\% disappearance in $\bar{\nu}_e$. 
Therefore, taking the aforementioned 2\% theoretical error on the flux prediction, point 1 could in principle explain the reactor anomaly, point 2 would predict too little disappearance, 
while point 3 would predict a slightly larger suppression of the flux. If the theoretical error were taken to be larger, for instance 
5\%~\cite{Hayes:2016qnu}, then all three points would be in agreement with the reactor data.

In a similar fashion, the Gallium anomaly is a discrepancy between the measured and theoretically predicted number of $\nu_e$ events in solar 
neutrino calibration experiments~\cite{Hampel:1997fc, Kaether:2010ag, Abdurashitov:1998ne, Abdurashitov:2005tb}. There,  $\nu_e$ is emitted by a  radioactive source and detected in a Gallium tank that contains the source. Although the $\nu_e$ flux is fairly well known, the detection cross section depends on nuclear physics form 
factors  with relatively large uncertainties~\cite{Bahcall:1997eg, Frekers:2011zz}. The ratio between measured and expected number of events is 
$R=0.84^{+0.054}_{-0.051}$~\cite{Kopp:2013vaa}. Thus, points 1 and 2 are consistent  with the Gallium anomaly within 2$\sigma$, while point 3 
would provide a better fit to these experiments.

\subsubsection{LSND and MiniBooNE}
\label{LSND}
\begin{figure}[!ht]
\centering
\includegraphics[width=0.42\textwidth]{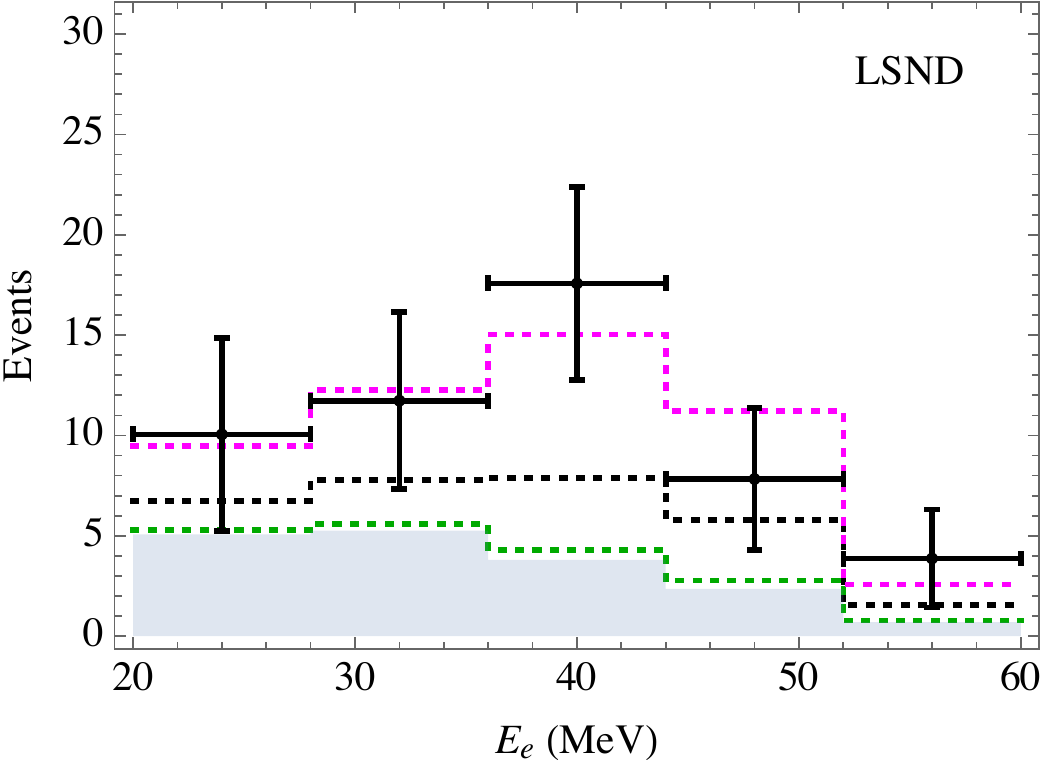}
\includegraphics[width=0.42\textwidth]{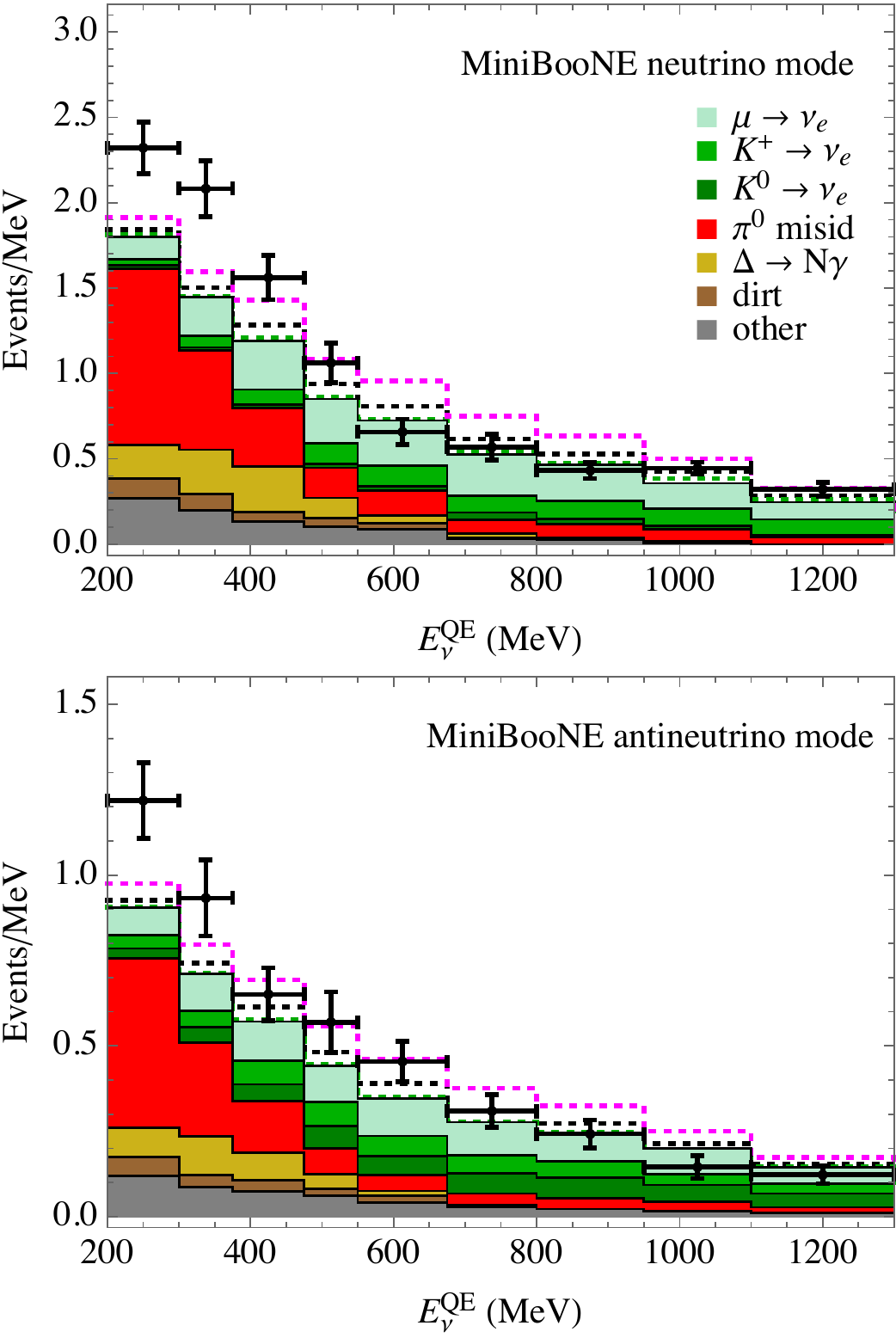}
\caption{\label{fig:miniboone} $\bar\nu_e$ appearance spectrum at LSND (top), and $\nu_e$ appearance MiniBooNE for the neutrino (middle) and antineutrino (bottom) modes.
The shaded histograms are the different background components as indicated in the legend (taken from Ref.~\cite{Aguilar:2001ty} for LSND and Ref.~\cite{Aguilar-Arevalo:2012fmn} for MiniBooNE). Black, green and magenta lines are for points 1, 2 and 3
in Table~\ref{t:points}, respectively.
}
\end{figure}

As we emphasized above, adding Dirac bulk mass terms splits the mass degeneracy between the three towers of KK modes and may lead to $\nu_\mu\to\nu_e$ appearance, thereby providing a possible explanation for the anomalies observed at LSND and MiniBooNE. We examined the event excess for the three benchmark points given in Table~\ref{t:points} in light of the full LSND~\cite{Athanassopoulos:1996ds, Aguilar:2001ty} and MiniBooNE data~\cite{Aguilar-Arevalo:2012fmn}, as shown in Fig.~\ref{fig:miniboone}. As we do not consider CP violation in the KK sector (we take all Yukawas and $c_i$ to be real), the appearance probabilities  $\nu_\mu\to\nu_e$ and $\bar\nu_\mu\to\bar\nu_e$ are nearly identical (apart from the small impact of matter effects). 

\begin{figure}
\centering
\includegraphics[width=0.45\textwidth]{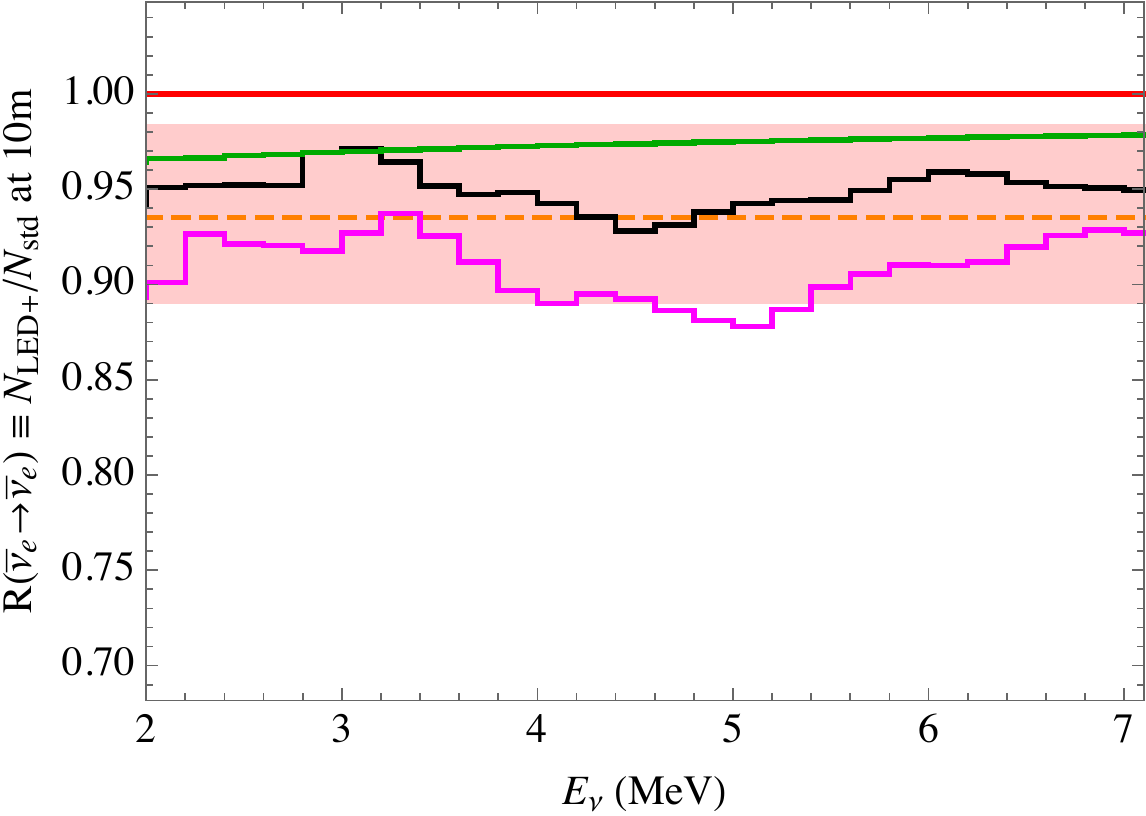}
\caption{Ratio between LED+ and standard oscillations for  $\bar\nu_e\to\bar\nu_e$ disappearance events at a reactor neutrino experiment for an illustrative baseline of 10~m. The orange dashed line is the ratio between the total number of observed to expected events. The light red band shows a 5\% theoretical uncertainty. The red, black, green and magenta lines display  the ratios for the standard scenario, and benchmark points 1,  2 and 3
in Table~\ref{t:points}, respectively\label{fig:reactor}.}
\end{figure}

For LSND (upper panel), the background is shown as the shaded histogram. We see clearly that point 3 (magenta line) could explain the excess quite well, while  point 1 (black line) gives rise to a smaller excess and point 2 (green line) essentially predicts no excess at all. We will discuss the impact of LED+ on MiniBooNE in a bit more detail.\footnote{Our simulation of MiniBooNE is more reliable than the LSND one, as we follow closely the MiniBooNE official data release, where the neutrino energy reconstruction comes from an official Montecarlo simulation. No similar information is available for LSND.}
For the neutrino mode, points 1 (black), 2 (green) and 3 (magenta) yield approximately 65, 8, and 176 excess events, respectively, in the region with neutrino energy from $200~\unit{MeV}$ to $1200~\unit{MeV}$.
For the antineutrino mode, the  excess events are 33, 4, and 89, respectively. Point 2 predicts very little excess due the small active neutrino mixing with the KK modes and typically small $\Delta m^2_{KK}$ for the relevant modes, as can be seen in Table~\ref{t:points}.
Moreover, we notice that the excesses, for points 1 and 3,  are found in the higher energy region, $E_\nu\sim400-800$~GeV. This is due to the $\Delta m^2_{KK}$ which is typically at the eV$^2$ scale or larger.
As a final comment, notice that to explain the LSND/MiniBooNE anomaly (point 3), we are slightly off the disappearance data as mentioned previously (see the magenta lines in Figs.~\ref{fig:minos} and \ref{fig:reactor}). We would like to point out that this tension is a common feature of sterile neutrino models which try to address the LSND/MiniBooNE anomalies~\cite{Kopp:2013vaa, Collin:2016aqd, Gariazzo:2017fdh}. This originates in the fact that $\nu_e$ ($\nu_\mu$) disappearance depends on $4|U_{e4}|^2$ ($4|U_{\mu4}|^2$) while the $\nu_\mu\to\nu_e$ appearance probability depends on $4|U_{e4}U_{\mu4}|^2$, and thus a non-zero appearance excess necessarily implies relevant $\nu_e$ and $\nu_\mu$ disappearance as well.

\subsection{Future oscillation experiments}
\label{futuretest}

\begin{figure}
\centering
\includegraphics[width=0.45\textwidth]{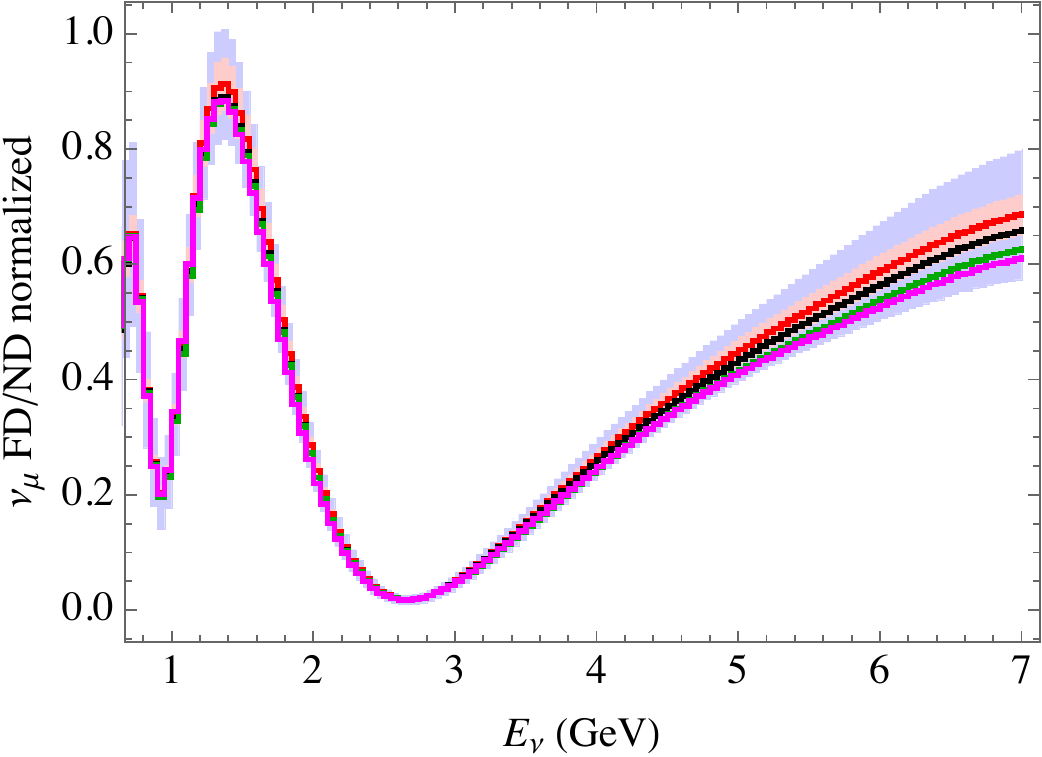}
\caption{\label{fig:dune} Near-to-far ratio for the $\nu_\mu\to\nu_\mu$ disappearance channel at DUNE. The red line is the expected ratio for the standard 3-neutrino framework. The light red (blue) band shows the 5\% systematic (statistical) uncertainty
 assuming 3 years run. Black, green and magenta are the ratios for points 1,  2 and 3 in Table~\ref{t:points}, respectively. }
\end{figure}

\subsubsection{DUNE} 
The sensitivity of the Deep Underground Neutrino Experiment (DUNE) to LED models without bulk mass terms was estimated in Ref.~\cite{Berryman:2016szd}: extra dimensions with $R\gtrsim (0.6~{\rm eV})^{-1}$ could be probed by DUNE for a massless lightest active neutrino.
This sensitivity is similar to the current constraint coming from the MINOS experiment.
In Fig.~\ref{fig:dune}, we present the  near-to-far event ratio, normalized to 1 in the absence of neutrino oscillations, for the DUNE experiment, assuming similar far and near detector acceptances, for the three benchmark points in our scenario. In our simulation we have used the energy resolution  from Ref.~\cite{DeRomeri:2016qwo}, of about 7\% at high energies for $\nu_\mu$ charged current (CC) events, and the $\nu_\mu$ flux and CC cross section from Ref.~\cite{Alion:2016uaj}, together with a 5\% normalization uncertainty.~\footnote{Note that the DUNE sensitivity to LED in Ref.~\cite{Berryman:2016szd} was derived using the energy resolution from DUNE CDR of about $20\%/\sqrt{E/{\rm GeV}}$, which compared to Ref.~\cite{DeRomeri:2016qwo} is slightly more aggressive for $E_\nu>4$~GeV and much more conservative otherwise.}
We assume a running time of 3.5 years in neutrino mode and a detector of 40 kton fiducial mass.
As expected, since the oscillation phase varies slower, deviations from standard oscillations are more easily observed at the high energy tail of the spectrum.
Point 1 seems to be rather challenging to be tested at DUNE. Points 2 and 3 have the potential to be probed, but that requires a detailed statistical analysis. See for example the DUNE sensitivity to point 3 in Fig.~\ref{fig:bound}.
Nevertheless, as we shall discuss next, our model will likely be first probed by the Fermilab Short-Baseline Neutrino Program.

\subsubsection{Short-Baseline Neutrino Program}
\begin{figure}[!ht]
\centering
\begin{subfigure}
\centering
\includegraphics[width=0.45\textwidth]{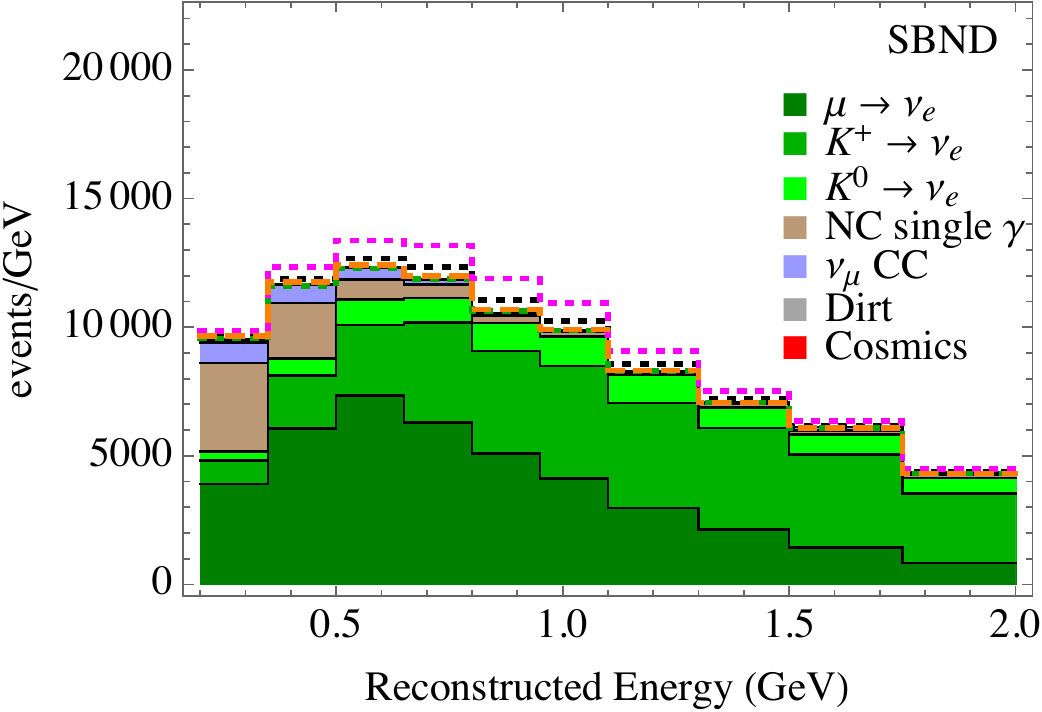}
\end{subfigure}
\begin{subfigure}
\centering
\includegraphics[width=0.45\textwidth]{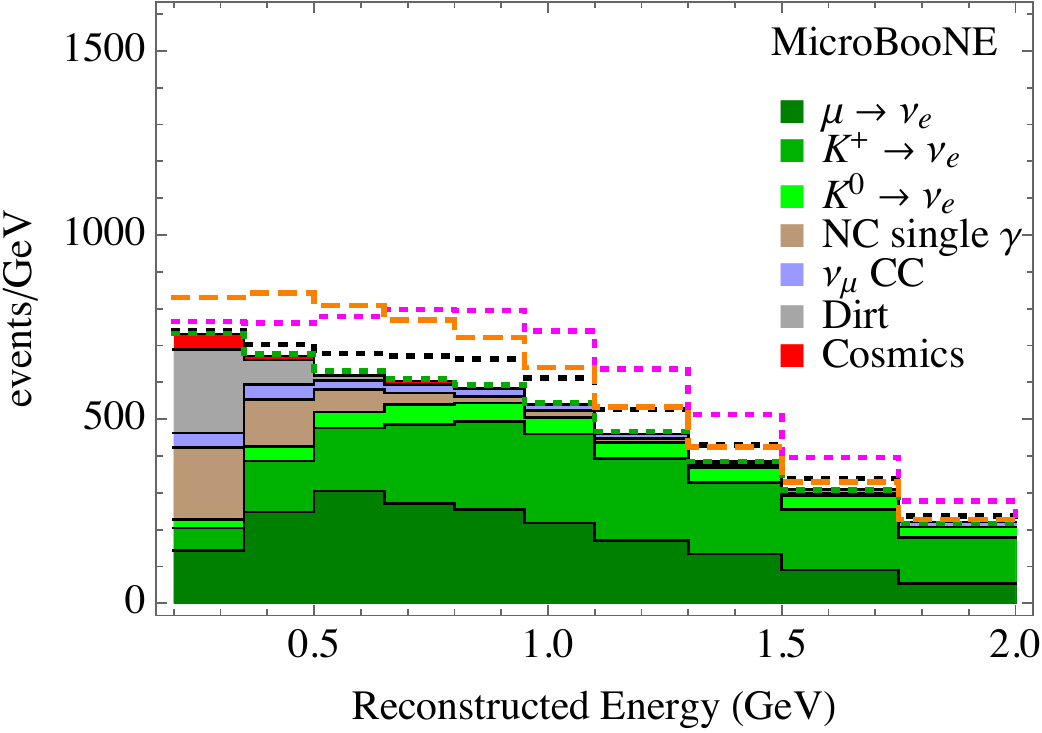}
\end{subfigure}
\begin{subfigure}
\centering
\includegraphics[width=0.45\textwidth]{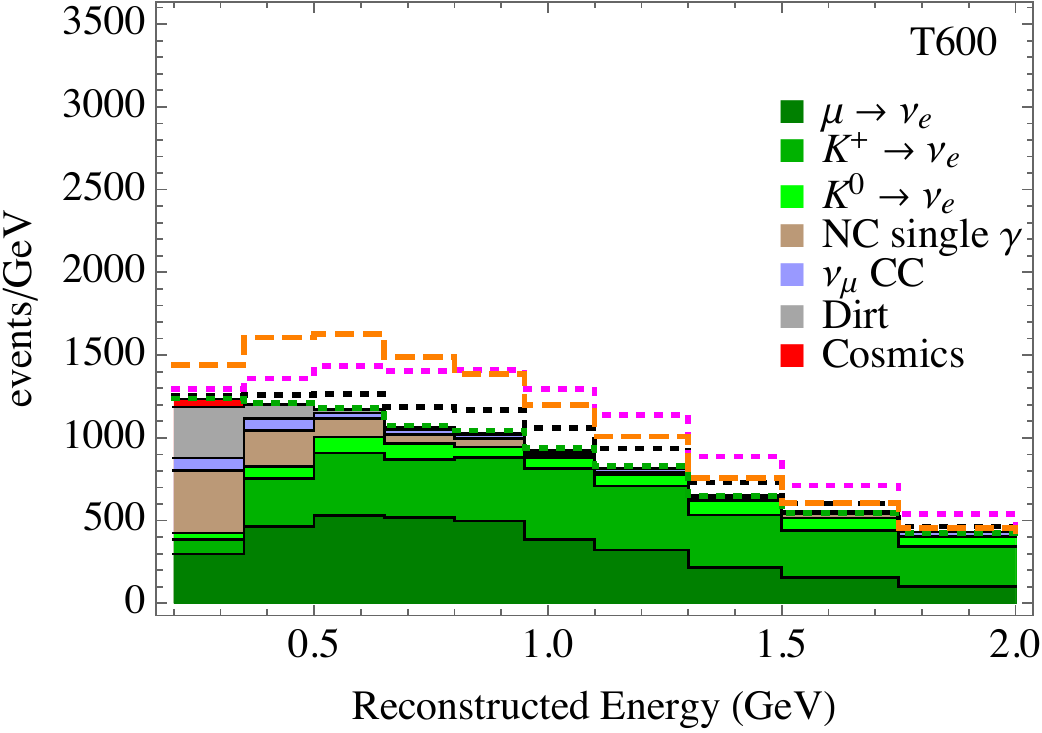}
\end{subfigure}
\caption{$\nu_e$ appearance spectrum at SBN detectors: (top) LAr1-ND, (middle) MicroBooNE, and (bottom) ICARUS-T600.
Full data set at these three detectors are assumed (see text). The shaded histograms are the different background components as indicated in the legend (taken from Ref.~\cite{Antonello:2015lea}). Black, green, magenta and orange lines are for points 1, 2, 3 
in Table~\ref{t:points} and the best-fit 3+1 sterile neutrino model, respectively.\label{fig:sbn}
}
\end{figure}

The Short-Baseline Neutrino Program (SBN) at Fermilab consists of three detectors, LAr1-ND, MicroBooNE and ICARUS-T600 with 
a distance from the target of 110~m, 470~m and 600~m, respectively. The MicroBooNE experiment has started to take data and will be 
able to investigate the source of the excess observed by MiniBooNE. 
Sterile neutrino models that explain the MiniBooNE excess do not necessarily affect the oscillation spectrum at LAr1-ND. However, in 
the LED+ framework, towers of KK modes would contribute to the oscillation spectrum at LAr1-ND. 
As a comparison, we show in Fig.~\ref{fig:sbn} the oscillation spectrum for the three benchmark points in LED+ and the best global fit 
point~\cite{Kopp:2013vaa} in a $3+1$ sterile neutrino model which has  $\Delta m_{41}^2 = 0.42$~eV$^2$, 
$\sin^2(2\theta_{\mu e})\equiv 4|U_{e4}U_{\mu4}|^2=0.013$. 
We used a luminosity of $6.6\times 10^{20}$,  $1.32\times 10^{21}$, and $6.6\times 10^{20}$ protons on target for LAr1-ND, MicroBooNE, and ICARUS-T600, respectively.
The event numbers predicted at the three detectors for our benchmark points and the best-fit $3+1$ sterile neutrino model are listed in Table~\ref{t:SBL}. By considering the spectra in Fig.~\ref{fig:sbn} and the event excesses in Table~\ref{t:SBL} at the three detectors, we observe that the SBN program at Fermilab has an excellent potential to probe LED+ models. This will be shown explicitly for a fixed set of Yukawas, $\lambda^i$, in the next section.

\begin{table}
\begin{center}
\begin{tabular}{l*{7}{c}r}
\hline
~~ & \multicolumn{2}{c}{{\bf SBND}}  &\multicolumn{2}{c}{{\bf MicroBooNE}} &\multicolumn{2}{c}{{\bf T600}}\\
\hline\hline
~~&Event&$S/\sqrt{B}$&Event&$S/\sqrt{B}$&Event&$S/\sqrt{B}$\\
\hline
Background&19800&&1070&&1940&\\
Point 1&498&3.5&93&2.84&169&3.8 \\
Point 2&54&0.38&11&0.34&21&0.47 \\
Point 3&1320&9.4&251&7.6&456&10.4\\
3+1 best fit &151&1.1&167&5.1&417&9.5\\
\hline
\end{tabular} 
\end{center}
\caption{\label{t:SBL}Event excesses and the predicted significances for the three detectors of the Short Baseline Neutrino Program for the various benchmark points we adopted, and the 3+1 best fit to the global  $\nu_\mu\to\nu_e$ appearance data~\cite{Kopp:2013vaa}. }
\end{table}

\subsection{Summary of oscillation constraints and sensitivities}
\begin{figure}
\includegraphics[width=0.45\textwidth]{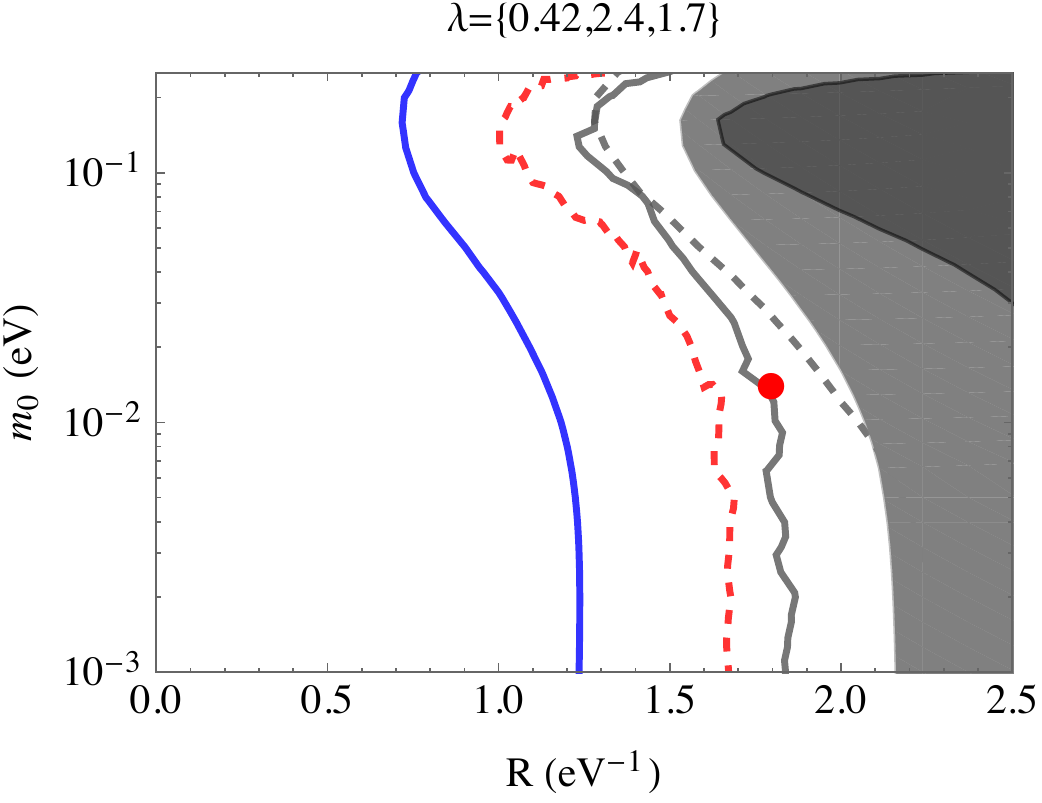}
\caption{
\label{fig:bound}
Present and future constraints, in the plane $R\times m_0$, by MINOS (gray solid line), reactor short baseline experiments (gray dashed line; 5\% flux uncertainty is assumed),  DUNE (red dashed line) and for the Short-Baseline Program at Fermilab (blue line; appearance mode only).
For reference, the light gray shaded region indicates large active-sterile mixing, namely $\sum_i(1-|W_i^{00}|^2)=0.3$. In the dark gray shaded region the approximation used to obtain $\Delta m^2_{21}$ and $\Delta m^2_{31}$ fails.
}
\end{figure}

To summarize the phenomenology of the model, we illustrate present constraints  and future sensitivities on LED+ in Fig.~\ref{fig:bound}.
For given values of $R$, the lightest active neutrino masses $ m_0$ and, as an example, a fixed set of $\lambda^i = (0.42,\,2.4,\,1.7)$ we calculate the values of $c_i$ in order to obtain the solar and atmospheric squared mass splittings. To perform this calculation we approximated the active neutrino masses using perturbation theory and Eqs.~(\ref{eq:mass-i}) and (\ref{eq:yuks}).
We present the estimated allowed region at $2\sigma$ level to the left of the corresponding line by the MINOS experiment (gray solid line) and 5\% deviation from the ratio of the observed event numbers to the SM prediction at reactor short baseline experiments (grey dashed line) in the plane $R - m_0$; as well as the projected $2\sigma$ sensitivities for the DUNE experiment (red dashed line) and for the Short-Baseline Program at Fermilab (blue line). For the latter, we used only the appearance channel. As a reference, our benchmark point 3 is shown with the red dot.
We also indicate the region in which the active-sterile mixing is large (light gray shaded), parametrized by $\sum_i(1-|W_i^{00}|^2)\ge 0.3$; as well as the region  in which our approximation for the evaluation of the solar and atmospheric mass splittings is not valid (dark gray shaded region). Fig.~\ref{fig:bound} highlights the potential of the near future neutrino Fermilab program in probing LED+ models. 

\subsection{Kinematic Constraints}
\label{sec:kin-constraints}
The limit on the  effective electron neutrino mass  from the Mainz experiment on Tritium decay is given by \cite{Kraus:2012he},
\begin{equation}
m_\beta \equiv\sqrt{\sum_{i=1}^3 |U_{ei}|^2 m_i^2} < \unit[2.3]{eV},
\end{equation}
for three active neutrinos. This bound comes from the analysis of the last $\unit[70]{eV}$ below the endpoint energy $Q= 18.572$~keV of the Tritium $\beta$ spectrum.  In our case, since we have KK modes heavier than $\unit[1]{eV}$, the above approximation fails (see e.g. Ref.~\cite{Esmaili:2012vg}) and the electron $\beta$ decay spectrum needs to be calculated exactly. 
This spectrum is given by
\begin{align}\label{eq:beta}
\beta(K_e, Q, |U_{ei}|, |m_\nu|) &= N_s F(Z,K_e)E_e p_e \times \\
& \times\sum_{i,j} P_i \xi_i |U_{e j}|^2\sqrt{\xi_i^2 - m_j^2}\, \mathcal{\theta} (\xi_i- m_j)\nonumber
\end{align}
\noindent where $K_e, p_e$ and $E_e$ are the electron kinetic energy, momentum and total energy, respectively. $F(Z, K_e)$ is the Fermi function and we will approximate it as a constant for a non-relativistic electron~\cite{Simpson:1981gq}. The energy window $\xi$ is defined as $\xi_i = Q- w_i - K_e$, where $w_i$ and $P_i$ are the excitation energy and transition probability for the excited state $i$ of the daughter nucleus, respectively. The detailed evaluation of the electron spectrum is given in Appendix~\ref{app:kinematics}.

To estimate the sensitivity of the Mainz experiment to our model, we define the deviation from the standard 3-neutrino predicted rate of events $S_{SM}$, for normal ordering and massless $\nu_1$, as
\begin{equation}
\delta R = 1-\frac{S}{S_{SM}},
\end{equation}
where $S$ is the rate of events in our model. The results are presented in Table~\ref{t:katrin}. For comparison, we also show $\delta R$  for a 3+1 sterile neutrino model with $\Delta m^2_{e4}= \unit[10.2]{eV}^2$ and mixing angle $\sin^2\theta_{e4} = 0.50$ (labelled \textbf{Sterile I}), which is marginally constrained by the Mainz experiment~\cite{Kraus:2012he}. As follows from Table~\ref{t:katrin}, the three data points in our model are less constrained than \textbf{Sterile I} data point. 
Compared to point 1 and point 3, point 2  predicts smaller deviation from the Standard 3-active neutrino model prediction, due to their lighter KK modes and smaller active neutrino mixing with KK modes. 

\begin{table}
\begin{center}
\begin{tabular}{l*{6}{c}r}
\hline
\rule{0pt}{3ex}   
&\textbf{Sterile I}&\textbf{Sterile II}& $P_1$& $P_2$  & $P_3$ \\
\hline\hline
\rule{0pt}{3ex} 
$10^5\times\delta R$&156&1.40&80.8&19.3&118\\
\hline
\end{tabular} 
\end{center}
\caption{\label{t:katrin} Deviation $\delta R$ from the standard 3-neutrino predicted rate of events for normal ordering and massless $\nu_1$,  for the three benchmark points and for two reference points in a $3+1$ sterile neutrino model.
\textbf{Sterile I}, with $\Delta m^2_{41}= \unit[10.2]{eV}^2$ and $\sin^2\theta_{e4} = 0.50$, is constrained by the Mainz experiment, while \textbf{Sterile II}, with $\Delta m^2_{41}= \unit[10.2]{eV}^2$ and  $\sin^2\theta_{e4} = 0.0045$, will be constrained by KATRIN.}
\end{table}

The future KArlsruhe TRItium Neutrino (KATRIN) experiment~\cite{Angrik:2005ep} will significantly improve the Mainz experiment bounds. For instance, it can probe a $3+1$ sterile neutrino model with $\Delta m^2_{41} = \unit[10.2]{eV}^2$ and the mixing angle $\sin^2\theta_{e4} = 0.0045$~\cite{ Esmaili:2012vg}. We also include this reference point (\textbf{Sterile II}) in Table~\ref{t:katrin}.
We expect that KATRIN will be able to test the three benchmark points in our model and probe a significant region of the LED+ parameter space. It would be also interesting to study the effect of LED+ on the shape of beta spectrum which is discussed for LED model in~\cite{Rodejohann:2014eka} if KATRIN would cover the entire beta spectrum~\cite{Mertens:2015ila}. We leave this to a future work.

\section{Other constraints}
\label{constraints}
\subsection{Constraints from Higgs decay}
The decay of the Higgs boson $h$ into a single KK mode is suppressed by the effective Yukawa coupling.  However the total width is enhanced by summing over all KK modes, where the approximate number of modes, e.g. in a single extra dimension model, is given by  $N\sim M_h R$. We calculate the $h$ decay width to all possible KK modes in the following. Neglecting the kinematic factor, the partial decay width to KK modes coming from $d$ extra dimensions is
\begin{align}
\Gamma_h &\sim \sum_{i=1}^{3}\sum^{m_{n}^i< M_h}_{n=0}\frac{M_h}{16\pi} (Y_n^i)^2\\
&\sim \frac{M_h}{8\pi} \sum_{i=1}^3\frac{\lambda^2_i}{M^{d}_5 V_{d}}  \prod_{k=1}^{d} (M_h R_{k})\\
&\sim \frac{M_h}{8\pi} \sum_{i=1}^3\lambda^2_i \left(\frac{M_h}{\pi M_5}\right)^d
\end{align}
where $V_d$ is their volume. For $d=2$, $M_5 = 10^6~\unit{GeV}$, and $\lambda_i\sim\mathcal{O}(1)$ we obtain $\Gamma(h\to {\rm KK~modes)}\sim 10^{-7}~\unit{MeV}$.
Therefore the $h$ decay width to KK neutrinos will not put any bound in LED+.

\subsection{Constraints from nucleosynthesis and supernova}
The presence of light KK modes can have an important impact on cosmological observations. For instance, nucleosynthesis data prefer the number of fully thermalized light species to be $N_{\rm{eff}} < 4$ even after doubling the systematic uncertainties~\cite{Abazajian:2012ys}. In Ref.~\cite{Barbieri:2000mg}, for LED without bulk masses and in the approximation of no matter asymmetry, it is shown none of the KK neutrinos are in thermal equilibrium with the plasma at MeV temperatures. The reason is that the matter effect induced by the plasma suppresses the mixing angles which are already small. In LED+ models, these  mixings are even smaller for the light KK modes, which may help evade nucleosynthesis bounds.
Summing over the energy density stored in all of these out-of-equilibrium 
KK neutrinos has large uncertainties 
and more work needs to be done to conclude
whether BBN data will constrain the parameter space of LED+  with interesting neutrino phenomenology.

In addition to cosmological bounds, astrophysical processes may be affected by the presence of light sterile neutrinos. One well known example is supernova explosion. In particular, SN1987a~\cite{Hirata:1987hu} is  likely to put constraints on $R$ since KK modes may carry away too much energy in the invisible channels from the supernova, thus modifying  its evolution~\cite{Barbieri:2000mg}. However, non-linear effects like collective neutrino oscillations~\cite{Cacciapaglia:2003dx} are still not well understood, and thus no robust bound can be derived on LED+ from these considerations. 

\section{Conclusions}
\label{conclusion}

In this article we have studied the properties of sterile neutrinos propagating in large extra dimensions with a bulk mass term of the order of $1/R$.  
By adding bulk masses to the standard LED scenario, the pattern of KK sterile neutrino masses and mixings can be significantly distorted. 
While in LED models the first KK mode dominates the oscillation phenomenology, and one can approximate the LED model by a specific 3+1 scenario, in LED+ models the mixings with the first KK modes can be suppressed by the bulk mass terms. This increases the relative importance of the higher KK modes and leads to distinct oscillation signatures. In LED+, the correspondence with a 3+1 scenario is lost: a large number of KK modes needs to be considered in order to obtain a reasonable approximation of the oscillation probability.

We have shown that the LED+ framework provides a well defined and testable scenario that has relevant implications for  neutrino oscillation experiments.  It has the potential to address the observed anomalies in short baseline neutrino experiments, namely the LSND/MiniBooNE anomalous  $\nu_\mu\to\nu_e$ appearance spectra, as well as the reactor and Gallium $\nu_e$ disappearance anomalies. We expect that the LED+ framework will be tested at the 
Short-Baseline Neutrino Program at Fermilab, and may also have an impact on the DUNE experiment, that may
then provide additional evidence for such scenario. Moreover, the KATRIN experiment will be able to probe a significant region of the LED+ parameter space.

\section{Acknowledgments}
We would like Thomas Carroll, Pilar Coloma, Andr\'{e} de Gouv\^{e}a, Enrique Fernandez-Martinez, Ornella Palamara, Zarko Pavlovic, Simon de Rijck and Zahra Tabrizi for useful discussions.
This manuscript has been authored by Fermi Research Alliance, LLC under Contract No. DE-AC02-07CH11359 with the U.S. Department of Energy, Office of Science, Office of High Energy Physics. The United States Government retains and the publisher, by accepting the article for publication, acknowledges that the United States Government retains a non-exclusive, paid-up, irrevocable, world-wide license to publish or reproduce the published form of this manuscript, or allow others to do so, for United States Government purposes.

Work at University of Chicago is supported in part by U.S. Department of Energy grant number DE-FG02-13ER41958. Work at ANL is supported in part by the U.S. Department of Energy under Contract No. DE-AC02-06CH11357. 
C.S.M. is supported by the S\~ao Paulo Research Foundation (FAPESP) under grant 2012/21627-9 and acknowledges the hospitality and support of the Fermilab theory group. YYL is supported by the Hong Kong
PhD Fellowship Scheme (HKPFS) and the Overseas Research Awards from Hong Kong University of Science and Technology. YYL would like to thank the hospitality of Enrico Fermi Institute, the University of Chicago and Fermilab where most of the work
was done.   
This project has also received partial funding from the European Union's Horizon 2020 
research and innovation programme under the Marie Sklodowska-Curie grant agreement No 674896 and No 690575. The work  of M.C. and C.W. was partially performed at the Aspen Center for Physics, which is supported by National Science Foundation grant PHY-1607611.

\appendix
\section{Fermions in the Linear Dilaton metric}
\label{fermion}
The Linear Dilaton (LD) metric is given by,
\begin{equation}
ds^2 =e^{\frac{2k|z|}{3}}(\eta_{\mu\nu} dx^\mu dx^\nu - dz^2),
\label{eq:metric}
\end{equation}
where we use the mostly-plus convention for the flat metric $\eta_{\mu\nu}$ and we assume $k>0$ (as argued in Ref.~\cite{Giudice:2016yja}, negative $k$ are equivalent to positive $k$ by coordinate transformations).

The dimensional deconstruction of the 5D model (with a bulk mass term) based on this metric leads to the  ``clockwork mechanism'' \cite{Giudice:2016yja,Kaplan:2015fuy} which was explored in several applications e.g. \cite{Hambye:2016qkf,Farina:2016tgd,Kehagias:2016kzt}.  
In the following, we are going to show that from the 5D perspective, fermions in the LD metric with a bulk mass can be put in equivalence with fermions in the LED with a bulk mass (see also \cite{Hambye:2016qkf}). The IR brane containing the SM fields is put at $z_0=0$ and the UV brane at $z_f= \pi R$. The 4D Planck scale $M_{\text{Pl}}$ is related to the 5D Planck scale $M_5$ by the following equation:
\begin{align}
M^2_{\text{Pl}} = 2 \int_{0}^{ \pi R} dz\,e^{k z} M_5^3  = \frac{2 M_5^3}{k}(e^{k \pi R}-1).
\end{align} 
Notice that the relation between $M_{\text{Pl}}$ and $M_5$ in LD is not equivalent to LED, see Eq.~(\ref{eq:Mpl}). 
Let us consider a fermion $\psi$ with a bulk mass term $M(z)$,
\begin{align}
\label{eq:action}
S_f = \int d^4 x \,dz\,\sqrt{g} \left[e
_A^Mi\bar{\Psi}_0\Gamma^A \overset{\leftrightarrow}{\partial_{M}}\Psi_0- M(z) \bar{\Psi}_0\Psi_0\right],
\end{align}
with the vierbein $e_A^M=e^{-\frac{1}{3} k z}\delta_A^M$ and $\Gamma^A = (\gamma^\mu,  i \gamma^5)$. The spin connection has been dropped since its contribution cancels in our case~\cite{Csaki:2007ns}. To satisfy the $S^1/Z_2$ symmetry, $M(z)$ must be odd under reflection: $M(z) = -M(-z)$. In addition, to canonically normalize the kinetic term, we use the field redefinition
\begin{equation}
\Psi_0 = e^{-\frac{2kz}{3}}\Psi.
\end{equation}
The action is then written as,
\begin{align}
\label{e:actionLD}
S_f = \!\int\!\! d^4 x \!\!\int^{\pi R}_{0} \!\!dz\left[i\bar{\Psi}\gamma^\mu\overset{\leftrightarrow}{\partial_{\mu}}\Psi - \bar{\Psi}\gamma_5\overset{\leftrightarrow}{\partial_{z}}\Psi - e^{\frac{k z}{3}} M(z)\bar{\Psi}\Psi\right].
\end{align}
If $M(z)=0$, the KK spectrum and wave functions are the same as those in LED without bulk mass. 
For the zero mode, this wavefunction is flat. In the LD metric we can see that the curvature does not help to localize this mode.

If we assume a non-zero bulk mass $M(z)$ for the LED or LD metrics, we can get that the zero mode is non-flat in both cases. In particular, if we identify $M(z)e^{\frac{k z}{3}}$ in Eq.~(\ref{e:actionLD}) with $c$ in Eq.~(\ref{eq:action-singlet}), we can conclude the equivalence between LED+  and LD with a corresponding bulk mass. In the case of gravitons it is possible to show that in the curvature in LD works as a mass term \cite{Antoniadis:2011qw} which reinforces the particular behavior of this metric. As a final comment, by having a small value for $k$ (about eV scale), one could obtain $M_5=10^6$~GeV for $R\sim\mathcal{O}(1/{\rm eV})$. This small $k$ is not unnatural as it is protected by the dilaton shift symmetry~\cite{Giudice:2016yja}.

\section{Minimal flavor violation realization}
\label{MFV}
In order to reduce the number of free parameters in our model and simplify the analysis, we assume a minimal flavor realization of the Yukawas and bulk mass terms as follows.
In the flavor basis, Eq.~(\ref{eq:action-singlet}) and the Yukawa terms in Eq.~(\ref{eq:Yukawa}) are written as
\begin{align}
S_f&=\int d^4x\,dz\,\left[i\bar{\Psi}_\alpha\Gamma^A \overset{\leftrightarrow}{\partial_{A}}\Psi_\alpha-\mathcal{C}_{\alpha\beta} \bar{\Psi}_\alpha\Psi_\beta\right],\nonumber \\
S_Y&=-\int d^4x \sum_{i=1}^3\left(\frac{\mathcal{Y}_{\alpha\beta}}{\sqrt{M_5}}\bar{L}_\alpha \tilde{H} \Psi^{R}_\beta (x^{\mu}, 0) + h.c.\right).
\end{align}
To have both $\mathcal{C}_{\alpha\beta}$ and $\mathcal{Y}_{\alpha\beta}$ diagonalized simultaneously in the ``intermediate basis" by the rotation in Eq.~(\ref{eq:rotate1}), we consider $\mathcal{C}_{\alpha\beta}$ to be  a polynomial of $\mathcal{Y}^{\dagger}\mathcal{Y}$,
\begin{equation}
\mathcal{C}_{\alpha\beta} = \left[\sum_a \mathcal{M}_a (\mathcal{Y}^{\dagger}\mathcal{Y})^{a}\right]_{\alpha\beta},
\end{equation}
where $\mathcal{M}_a$ are dimensionful coefficients. Then we have
\begin{equation}
c_i = \sum_a \mathcal{M}_a (\lambda^2_i )^{a}.
\end{equation}
Different values of $\mathcal{M}_a$ were chosen to get the parameters we used in 
our simulation. For example, we have set
\begin{align}
\mathcal{M}_0 = -0.44~\unit{eV}, ~~ \mathcal{M}_1 = 3.9~\unit{eV}, \nonumber\\
\mathcal{M}_2=-2.6~\unit{eV}, \mathcal{M}_{a>2} = 0~\unit{eV}
\end{align}
to get the bulk mass values used for point 2. Without loss of generality we have neglected higher order contributions to $c_i$. 

 \section{Details for kinematic constraints}
 \label{app:kinematics}
We present here some details of the beta decay rate calculation used in Section~\ref{sec:kin-constraints}.
Using Eq.~({\ref{eq:beta}), the rate $S$ of the electrons passing the potential barrier $Uq$ and arriving at the detector is
given by
\begin{equation}
S = \int_0^\infty\beta(K_e, Q, |U_{ei}|, |m_\nu|) T(K_e, qU) dK_e,
\end{equation}
where $T(K_e, qU)$ is the transmission function. We  use the following approximation to get a conservative estimate of $S$,
\begin{eqnarray}
T(K_e, qU) = \begin{cases}
                     1~~~ {\rm if} ~~~K_e > qU \\
                     0~~~{\rm if} ~~~K_e < qU
                     \end{cases}\,.
\end{eqnarray}
In addition, we approximate $\beta(K_e, Q, |U_{ei}|, |m_\nu|)$ as follows:
\begin{align}
\beta(K_e, Q, &|U_{ei}|, |m_\nu|) \simeq C \sqrt{(Q-\xi)} \xi  \times\\
&\times\sum_{j=1}^3 \sum_{n=0}^\infty |U_{ej}|^2 |W^{0n}_j |^2 \sqrt{\xi^2 - m_{j,n}^2} \,\theta(\xi-m_{j,n}),\nonumber
\end{align}
where we assumed that the daughter nucleus is in the ground state with $\xi = Q-K_e$ and we used the non-relativistic relation for electron momentum, $p_e\simeq\sqrt{2m_e K_e}$ ($m_e$ is the electron mass). $C\simeq N_s F(Z, K_e) E_e \sqrt{2m_e}$ is approximately a constant. We obtain the expected event rate $S$ by integrating over $\unit[1]{eV} <\xi < \unit[70]{eV}$, which is equivalent to set the potential barrier $q U = Q-70~\unit{eV}$.

\bibliography{LEDBulk_neutrino}
\end{document}